\documentclass{aa}
\usepackage{natbib}
\usepackage{graphicx}
\usepackage{txfonts}
\usepackage{hyperref}
\usepackage{booktabs}
\newcommand{\numax}{\mbox{$\nu_{\rm max}$}\xspace}
\newcommand{\deltanu}{\mbox{$\langle \Delta\nu \rangle$}\xspace}
\newcommand{\teff}{\mbox{$T_{\rm eff}$}\xspace}
\newcommand{\logg}{\mbox{$\log g$}\xspace}
\newcommand{\feh}{\mbox{$\rm{[Fe/H]}$}\xspace}

\newcommand{\afe}{\mbox{$\rm{[\alpha/Fe]}$}\xspace}

\newcommand{\muas}{\mbox{$\mu \rm as$}\xspace}

\newcommand{\kepler}{\emph{Kepler}\xspace}
\newcommand{\gaia}{\emph{Gaia}\xspace}
\newcommand{\ktwo}{K2\xspace}
\newcommand{\tess}{TESS\xspace}

\defcitealias{Lindegren2021}{L21}
\defcitealias{Khan2019}{K19}
\defcitealias{Mosser2011a}{M11}
\defcitealias{Mosser2009}{MA09}
\defcitealias{Elsworth2020}{E20}
\defcitealias{Yu2018}{Y18}

\usepackage[inline]{enumitem}
\usepackage{gensymb}
\usepackage{placeins}
\usepackage{subcaption}

\begin{document}
  \title{Investigating \gaia EDR3 parallax systematics using asteroseismology of Cool Giant Stars observed by \\ \kepler, \ktwo, and \tess}
  \subtitle{II. Deciphering \gaia parallax systematics using red clump stars}
  \titlerunning{Deciphering \gaia parallax systematics using red clump stars}

  %\subtitle{}
  \author{S. Khan\inst{1}
        \and
        R. I. Anderson\inst{1}
        \and
        A. Miglio\inst{2,3,4}
        \and
        B. Mosser\inst{5}
        \and
        Y. P. Elsworth\inst{4}
                }

  \institute{Institute of Physics, \'Ecole Polytechnique F\'ed\'erale de Lausanne (EPFL), Observatoire de Sauverny, 1290 Versoix, Switzerland\\
              \email{saniya.khan@epfl.ch}
         \and
             Dipartimento di Fisica e Astronomia, Università degli Studi di Bologna, Via Gobetti 93/2, I-40129 Bologna, Italy
         \and
             INAF - Osservatorio di Astrofisica e Scienza dello Spazio di Bologna, Via Gobetti 93/3, I-40129 Bologna, Italy
         \and
         	 School of Physics and Astronomy, University of Birmingham,
         	Edgbaston, Birmingham, B15 2TT, UK
         \and
         	 LESIA, Observatoire de Paris, PSL Research University, CNRS, Sorbonne Universit\'e, Universit\'e Paris Diderot, 92195 Meudon, France
            }

  \date{Received MM DD, YYYY; accepted MM DD, YYYY}

  \abstract
	{We analyse \gaia EDR3 parallax systematics as a function of magnitude and sky location using a recently published catalogue of 12,500 asteroseismic red-giant star distances. We selected $\sim 3500$ red clump (RC) stars of similar chemical composition as the optimal subsample for this purpose because 1) their similar luminosity allows to straightforwardly interpret trends with apparent magnitude; 2) RC stars are the most distant stars in our sample at a given apparent magnitude so that uncertainties related to asteroseismic radii/distances are smallest; 3) they provide the largest sample of intrinsically similar stars. We perform a detailed assessment of systematic uncertainties relevant for parallax offset estimation based on the asteroseismic distances. Specifically, we investigate a) the impact of measuring the basic asteroseismic quantities \numax and \deltanu using different pipelines, b) uncertainties related to extinction, c) the impact of adopting spectroscopic information from different surveys, and d) blending issues related to photometry. Following this assessment, we adopt for our baseline analysis the asteroseismic parameters measured as in \citet{Elsworth2020}, spectroscopy from the Apache Point Observatory Galactic Evolution Experiment (DR17), and we further restrict the sample to low-extinction ($A_V \le 0.5$ mag) RC stars with quality astrometric solutions from \gaia EDR3 as indicated by RUWE $< 1.4$. We then investigated both the parallax offset relative to the published \gaia EDR3 parallaxes and the residual parallax offset after correcting \gaia EDR3 parallaxes following \citet{Lindegren2021}. 
	
	We find residual parallax offsets very close to zero ($-1.6 \pm 0.5$ (stat.) $\pm 10$ (syst.) \muas) for stars fainter than $G > 11$ mag in the initial \kepler field, suggesting that the Lindegren parallax offset corrections are adequate in this magnitude range. For 17 \ktwo campaigns in the same magnitude range, the residual parallax offset is $+16.5 \pm 1.7$ (stat.) $\pm 10$ (syst.) \muas. At brighter magnitudes ($G \le 11$ mag), we find inconsistent residual parallax offsets between the \kepler field, 17 \ktwo campaigns, and the \tess southern continuous viewing zone, with differences of up to $60$ \muas. This contradicts studies suggesting a monotonic trend between magnitude and residual parallax offsets and instead suggests a significant dependence on sky location at bright magnitudes due to the lack of bright physical pairs available for determining the parallax offset corrections. Inspection of the 17 \ktwo campaigns allows to investigate parallax offsets as a function of ecliptic longitude and reveals a possible signal. Finally, we estimate the absolute magnitude of the RC and obtain $M_{K_s}^{\rm RC} = -1.650 \pm 0.025$ mag in the 2MASS $K_s$-band and $M_{G}^{\rm RC} = (0.432 \pm 0.004) - (0.821 \pm 0.033) \cdot (\teff [\rm{K}] - 4800\rm{K})/1000\rm{K}$ [mag] in the \gaia $G$-band.}



\maketitle

%________________________________________________________________

\section{Introduction}
\label{sec:intro}

\gaia EDR3 has provided unprecedented data that generate a lot of interest in the astrophysical community. It is also known that there exists a residual parallax offset at the level of $\sim 10 \, \muas$ that is an issue for distances beyond 1 kpc \citep{GaiaCollaboration2021}. \citet[][hereafter \citetalias{Lindegren2021}]{Lindegren2021} estimated the parallax bias as a function of magnitude, colour, and ecliptic latitude. The bias corrections by \citetalias{Lindegren2021} were determined using quasars, Large Magellanic Cloud (LMC) sources, and physical pairs. 

Many studies have reported parallax offsets ($\Delta \varpi = \varpi_{\rm EDR3} - \varpi_{\rm other}$) and residuals ($\Delta \varpi_{\rm corr} = (\varpi_{\rm EDR3} - Z_5) - \varpi_{\rm other}$) to complement the work done by \citetalias{Lindegren2021}, where $\varpi_{\rm other}$ is the parallax measured through an independent method and $Z_5$ is the \citetalias{Lindegren2021} offset for 5-parameter astrometric solutions. These methods are either based on direct comparisons where the parallaxes are known or measured independently (e.g. quasars, detached eclipsing binaries, asteroseismology), jointly determined as part of the calibration of period-luminosity relations (e.g. Cepheids, RR Lyrae), or through differential methods (e.g. binaries, open and globular clusters). We refer the reader to the introduction of Paper I \citep{Khan2023} and to Fig. \ref{fig:compilation} for a review of residual parallax offsets that have been measured in the literature. 
Existing compilations of literature results showed that parallax offset residuals potentially follow a trend with magnitude: at the brightest magnitudes ($6 < G < 13$ mag), \citetalias{Lindegren2021} corrections would overcorrect parallaxes by +10 to 20 $\rm \mu as$; while, as one moves to fainter magnitudes ($G > 13$ mag), the residuals are decreasing to about $\sim +5 \, \rm \mu as$ (see e.g. Fig. 1 in \citealt{Li2022}; Fig. 2 in \citealt{Riess2022}; Fig. 10 in \citealt{Molinaro2023}; where $\Delta \varpi_{\rm corr}$ is defined with the opposite sign). 

\begin{figure*}
	\centering
	\includegraphics[width=\hsize]{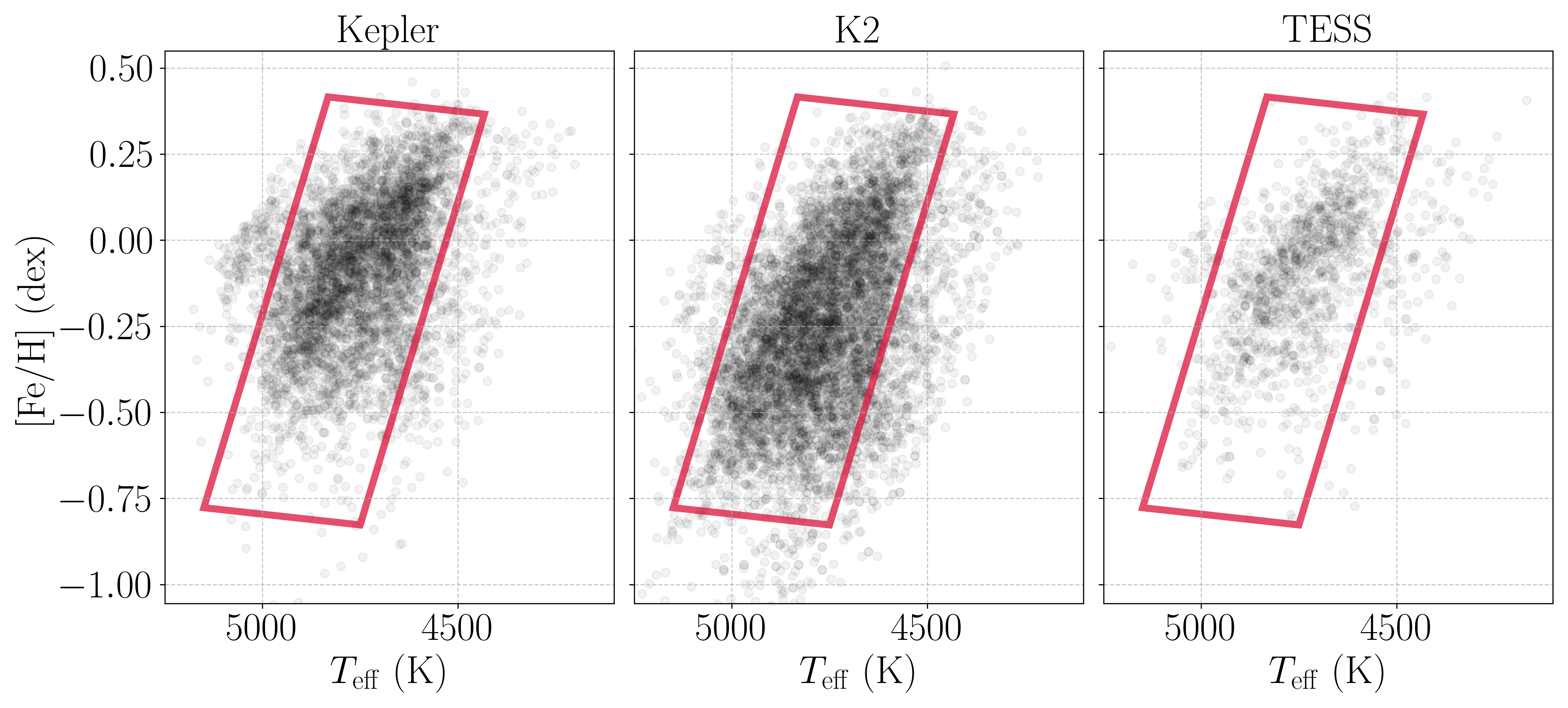}
	\includegraphics[width=\hsize]{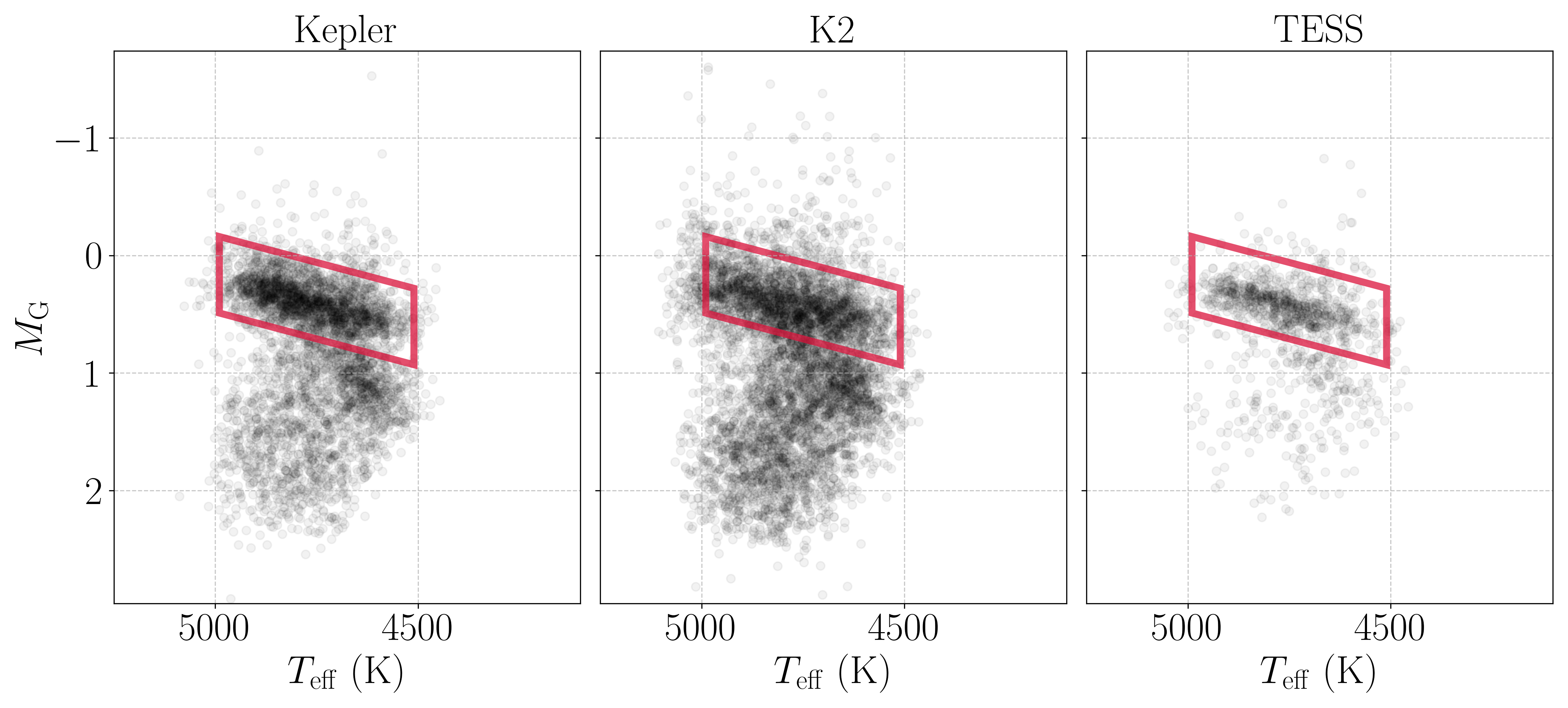}
	\caption{Diagrams illustrating our two-step selection method to have a RC population as homogeneous as possible among the \kepler (left), \ktwo (middle), and \tess-SCVZ fields (right), using \citetalias{Elsworth2020} and APOGEE DR17. \textit{Top:} \teff vs \feh. The red box indicates the first selection of the bulk of RGB and RC stars, to make sure that our sample is chemically similar. \textit{Bottom:} Hertzsprung-Russell Diagrams (\teff vs absolute magnitude from \gaia in the $G$-band), after the first selection has been applied. We draw a region in the HRD to select RC stars (red rectangles).}
	\label{fig:hrd}
\end{figure*}

In Paper I, we presented distance measurements for red-giant stars in the original \kepler mission \citep{Borucki2010}, 17 \ktwo campaigns \citep{Howell2014}, and the \tess southern continuous viewing zone \citep[\tess-SCVZ;][]{Ricker2015} fields, with a first comparison to \gaia parallaxes. \kepler and \ktwo correspond to the same telescope but with very different baselines (4yr and 80d, respectively). \tess is a smaller telescope with intermediate baseline in the continuous viewing zones (1yr). The asteroseismic distance is determined by assuming the luminosity of a stellar model that produces the observed asteroseismic observables. The current paper seeks to address specifically the best way to determine parallax systematics based on that dataset, and to determine the limitations of this approach based on asteroseismology.

We refer the reader to Paper I for the description of the datasets and of the methods to derive distances based on asteroseismic and spectroscopic constraints. Section \ref{sec:rc} discusses the key role of red clump stars as distance indicators. In Sect. \ref{sec:syst}, we explore and quantify systematics that may affect the estimation of the parallax offset, such as the asteroseismic method, the extinction, the spectroscopic survey, and the photometric bands. This allows us to compute a total systematic uncertainty on the parallax zero-point estimate, as well as define the most suitable sample to characterise the \gaia offset. Section \ref{sec:gaiazp} presents our final comparison of \gaia EDR3 and asteroseismic parallaxes, and compares the resulting offsets with those predicted by \citetalias{Lindegren2021}. We also discuss distinctions that need to be made between \kepler and \ktwo observations, trends that appear when we consider \ktwo campaigns individually, the distribution of bright calibrators in the \citetalias{Lindegren2021} model, as well as the possibility to derive the magnitude of the red clump based on \kepler results in Sect. \ref{sec:discussion}. Conclusions are given in Sect. \ref{sec:conclusions}.

\section{Red clump stars as key distance indicators}
\label{sec:rc}

After the helium flash event, low-mass stars settle in what is known as the red clump, where helium burning proceeds in the core. At this stage, stars share very similar helium core masses and hence luminosities on the horizontal branch. They are thus located in a confined region of the Hertzsprung-Russell Diagram (HRD), with a small dependence on effective temperature owing to their total mass (lower masses being slightly hotter) and composition (higher metallicities being cooler). For this reason, RC stars are known as standard candles, as their apparent brightness relates directly to their distance. They also find applications in e.g. estimating extinctions \citep[e.g.][]{Skowron2021,Sanders2022}, mapping the Galactic bulge \citep[e.g.][]{Paterson2020,Johnson2022}, and constraining stellar physics processes \citep[e.g.][]{Bossini2015,Bossini2017}. We refer the reader to \citet{Girardi2016} for an extensive review about the red clump phase.

Based on the HRD shown on Fig. \ref{fig:hrd}, we see that among the most luminous giants in our sample, RC stars are more common than first-ascent red-giant branch (RGB) stars. For a given apparent magnitude, it means that RC stars will in general be at further distances compared to RGB stars. This has at least two benefits for parallax offset estimates: 1) asteroseismic distances have a relative uncertainty ($\lesssim 5\%$) that will transform into a larger absolute error for nearby stars (larger parallaxes) compared to distant ones (smaller parallaxes; see Sect. 5.3 in Paper I); 2) the red clump alone is a rather homogeneous population compared to the whole sample of giants, which may help to reduce dispersion among the asteroseismic parallaxes. Some studies have shown that radii/masses of old metal-poor red horizontal branch (RHB) stars are likely to be biased \citep[see, e.g.,][]{Tailo2022,Matteuzzi2023}, but we can assert that very few RHB stars are included in our datasets (if any; see also Sect. \ref{sec:kepvsk2}). Besides, we remind the reader that our asteroseismic parallaxes are model-dependent. In this regard, homogeneity is important because it ensures that the systematics affecting our stars are similar throughout the sample. This allows us to obtain the most precise relative parallax differences (e.g. between asteroseismic pipelines in Sect. \ref{sec:syst_astero}), as a function of magnitude or sky position. 

We strive to have a RC selection as homogeneous as possible among our \kepler, \ktwo, and \tess-SCVZ datasets. These surveys cover different areas on the sky and so we want to limit the impact of sample differences as much as possible. As a first step, we select the bulk of RGB and RC stars in the \teff---\feh plane, with the following corner coordinates: \{(4832, 0.416), (4432, 0.366), (4750, $-$0.827), (5150, $-$0.777)\}. This ensures that our giant stars are chemically comparable, and that we do not include any RHB stars. Then, we select stars within the range defined by the following (\teff, $M_G$) coordinates: \{(4990, $-$0.160), (4510, 0.280), (4510, 0.930), (4990, 0.490)\} (see Fig. \ref{fig:hrd}), where $M_G$ is computed using \gaia parallaxes (without \citetalias{Lindegren2021} corrections) and extinctions $A_V$ from \texttt{PARAM}. Given the impact that a parallax offset would have on $M_G$ and the height of the box, it does not need to be considered at this stage of the analysis. In the \kepler field, core-helium burning stars have been identified using the evolutionary-dependent signature of gravity modes in the oscillation spectra \citep{Bedding2011,Elsworth2017}, hence we used this classification to prevent any contamination from RGB stars. For \ktwo and \tess, we know that there could still be some contamination due to RGB stars, as well as secondary clump stars \citep[see, e.g., Sect. 3.5 in][]{SchonhutStasik2023}. Some studies have provided classification for \ktwo campaigns (K2 GAP, \citealt{Zinn2022}; K2-APO, \citealt{SchonhutStasik2023}) and \tess continuous viewing zones \citep[HD-TESS,][]{Hon2022}, although not as robust as in \kepler's dataset. The original \kepler field remains unmatched in its asteroseismic detail due to the long temporal baseline of 4 years and the larger telescope aperture compared to \tess. The crossmatch between our RC stars and those catalogues leads to lower statistics and does not affect our results, hence we continued with our selection relying on \teff, \feh, and $M_G$. Without the RGB/RC classification for the \kepler field, we would find a contamination of about 15\% by RGB stars.

We therefore investigate parallax systematics using primarily RC stars. In total, we selected 3422 RC stars after cuts on $A_V$ and RUWE (see Table \ref{table:fields} and beginning of Sect. \ref{sec:gaiazp}). Where useful, we consider other RGB stars as well (Sects. \ref{sec:syst_k2} and \ref{sec:syst_spectro}). General information about the RC stars selected is provided in Table \ref{table:fields}.

\begin{table}
	\small
	\caption{Summary information for the RC stars selected in each field.}
	\label{table:fields}
	\centering
	\begin{tabular}{c | c c c c c c}
		\hline\hline
		Field & $(l,b)$ & $\langle A_V \rangle$ & $\langle \sigma_{\rm R}/R \rangle$ & $t_{\rm obs}$ & $N$ & $N'$ \\
		\hline
		\kepler & (76, 13) & 0.22 & 0.018 & 4 yrs & 1729 & 1560 \\
		\hline
		\ktwo & - & 0.28 & 0.038 & 80 d & 2230 & 1227 \\
		C1 & (264, 58) & 0.15 & 0.034 & - & 48 & 47 \\
		C2 & (354, 18) & 0.89 & 0.04 & - & 157 & 9 \\
		C3 & (51, -52)  & 0.16 & 0.038 & - & 126 & 118 \\
		C4 & (172, -26) & 0.85 & 0.037 & - & 482 & 38 \\
		C5 & (209, 31) & 0.14 & 0.039 & - & 291 & 271 \\
		C6 & (321, 50) & 0.18 & 0.036 & - & 170 & 159 \\
		C7 & (14, -15) & 0.45 & 0.038 & - & 145 & 77 \\
		C8 & (129, -57) & 0.15 & 0.04 & - & 112 & 104 \\
		C10 & (291, 58) & 0.09 & 0.042 & - & 43 & 37 \\
		C11 & (1.3, 7.2) & 1.23 & 0.035 & - & 83 & 0 \\
		C12 & (77, -60) & 0.14 & 0.037 & - & 112 & 107 \\
		C13 & (180, -15) & 1.24 & 0.037 & - & 183 & 0 \\
		C14 & (241, 53) & 0.11 & 0.037 & - & 88 & 83 \\
		C15 & (347, 28) & 0.74 & 0.031 & - & 4 & 0 \\
		C16 & (209, 35) & 0.13 & 0.037 & - & 102 & 95 \\
		C17 & (319, 54) & 0.13 & 0.042 & - & 73 & 71 \\
		C18 & (209, 31) & 0.14 & 0.045 & - & 11 & 11 \\
		\hline
		\tess-SCVZ & (276, -30) & 0.24 & 0.035 & 1 yr & 731 & 635 \\
		\hline
	\end{tabular}
\tablefoot{Galactic coordinates, mean extinction (in mag), mean relative uncertainty on radius, duration of observations, and number of stars before and after applying the cuts summarised in Sect. \ref{sec:gaiazp} ($A_V \leq 0.5$ mag and RUWE $< 1.4$). In terms of ecliptic latitude, \kepler is at $+65^{\circ}$, \ktwo at around $0^{\circ}$, and \tess at $-90^{\circ}$.}
\end{table}

\section{Quantifying systematic uncertainties related to asteroseismic parallaxes}
\label{sec:syst}

Asteroseismic parallaxes are estimated with the Bayesian tool \texttt{PARAM} \citep{Rodrigues2017}. The code requires the following observational parameters as inputs: \numax, \deltanu, \teff, \logg, \feh, and \afe (when available), as well as magnitudes in various photometric systems. Observations are then compared with a grid of stellar evolutionary tracks in order to predict the best-fitting stellar properties, such as radii, masses, distances, total extinctions, etc.

\subsection{Differences among asteroseismic methods}
\label{sec:syst_astero}

The global asteroseismic parameters that are used to determine the model luminosity can differ according to the way that they are measured. In our study, we have access to asteroseismic observables determined by \citet[][hereafter \citetalias{Elsworth2020}]{Elsworth2020} and \citet[][hereafter \citetalias{Mosser2009}]{Mosser2009}. We use RC stars to assess how using one or the other affects our results. 

As the relative uncertainty on the seismic distance is constant ($\lesssim 5\%$), we expect that the resulting absolute error on parallax will be larger for nearby stars (large parallaxes), compared to distant stars (small parallaxes; see also discussion in Sect. 5.3 in Paper I). Because of the very restricted range of absolute magnitudes in the RC, there is a tight relation between distance and apparent magnitude: nearby stars are bright, while more distant stars are fainter. 

\begin{figure}[!t]
	\centering
	\includegraphics[width=\hsize]{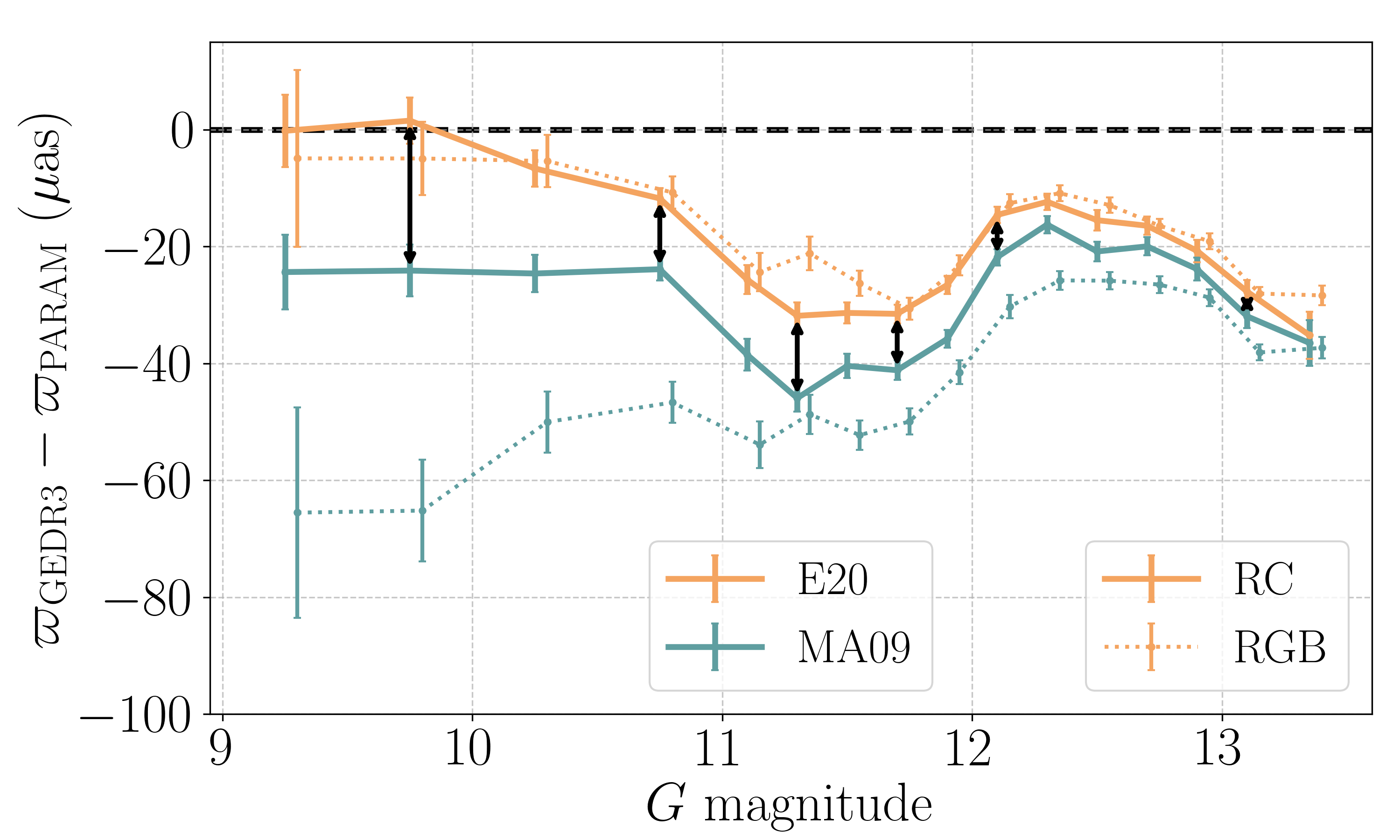}
	\caption{Parallax difference binned as a function of magnitude for the \kepler field, with RC (solid) and RGB stars (dotted line) shown separately. Results from \citetalias{Elsworth2020} and \citetalias{Mosser2009} are shown in orange and green, respectively. The black arrows show the difference used in order to measure the systematic due to using different asteroseismic methods.}
	\label{fig:astero}
\end{figure}

Figure \ref{fig:astero} shows that \citetalias{Elsworth2020}'s pipeline yields greater consistency between RGB and RC stars: differences are in the order of a few \muas at most. By contrast, with \citetalias{Mosser2009}'s asteroseismic constraints, discrepancies cover a wide range of values between 10 \muas and 40 \muas at the brightest magnitudes. Owing to the greater agreement based on the \citetalias{Elsworth2020} pipeline, we use their asteroseismic quantities for the remainder of this study. We adopt the difference of the two pipelines as the asteroseismic contribution to the total systematic uncertainty.

As illustrated in Fig. \ref{fig:astero}, we estimate the bias due to asteroseismic methods as the absolute mean difference between the parallax offsets measured using \citetalias{Elsworth2020} and \citetalias{Mosser2009}, also considering how it evolves as a function of magnitude. For \kepler, it decreases from 15 to 5 \muas; for \ktwo, the values are rather stable and stay around 5-10 \muas at most; and for \tess, it is around 8 \muas. It is clear that the bias is of greater importance at the brightest magnitudes, and decreases as we move towards fainter stars. This is also in accordance with the above discussion. The absolute parallax uncertainty of \gaia parallaxes is lower than asteroseismic parallax uncertainties at $G \lesssim 10$-$11$ mag for a fiducial RC star of $M_G=0.5$ mag with $\teff = 4750$ K. Overall, the difference between the \citetalias{Elsworth2020} and \citetalias{Mosser2009} parallax offsets, which we adopt as an asteroseismic systematic uncertainty, we measure can be approximated by:
\begin{align*}
	\sigma_{\rm seismo} \sim 14 + 6 \cdot (11 - G) \; \rm \muas. 
\end{align*}
This relation is valid for $G \in$ [9, 13] mag.

\subsection{Impact of extinction}
\label{sec:syst_extinction}

Extinction is an important quantity to consider in at least two aspects: its determination is often quite uncertain, and reddening is such that the observed colour is different from the star's intrinsic colour. In \texttt{PARAM}, extinction coefficients are computed adopting the \citet{Cardelli1989} and \citet{ODonnell1994} reddening laws with $R_V = 3.1$. It is then assumed that extinctions in all filters $A_{\lambda}$ are related by a single interstellar extinction curve expressed in terms of its $V$-band value, i.e. $A_{\lambda} (A_V)$. The total extinction $A_V$ and the distance $d$ can then be derived simultaneously. Hence, uncertainties come from both the reddening law (dust composition) and the match of the assumed, reddened SED to the observed SED.

\subsubsection{\ktwo fields: Galactic plane versus halo}
\label{sec:syst_k2}

\begin{figure}
	\centering
	\includegraphics[width=\hsize]{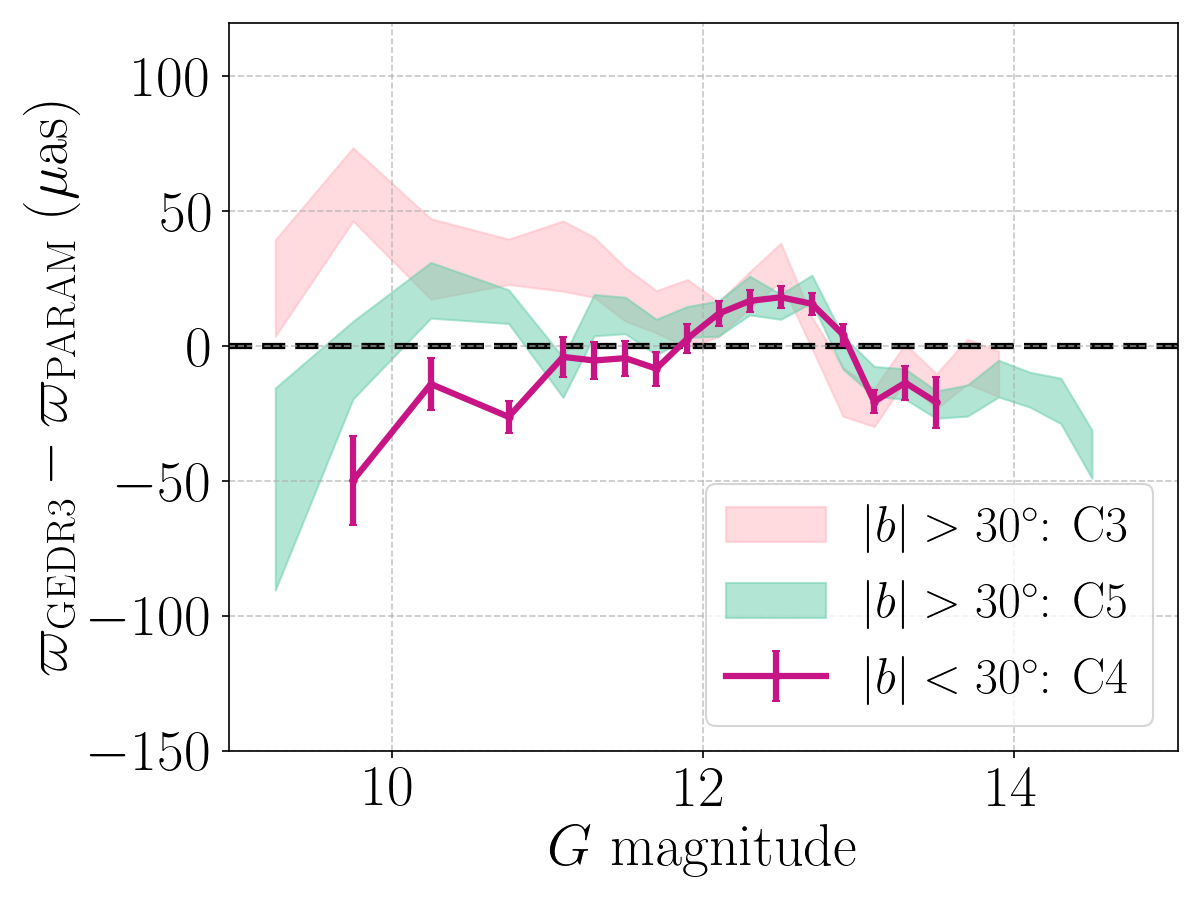}
	\caption{Parallax difference binned as a function of magnitude for three \ktwo campaigns, selected because of large number of stars and low/high extinction: C3 (light pink) and C5 (light green) which are located in the halo ($|b| > 30^\circ$), and C4 closer to the Galactic plane (magenta; $|b| < 30^\circ$).}
	\label{fig:k2fields}
\end{figure}

Figure \ref{fig:k2fields} shows the relation between the parallax difference and the $G$ magnitude, for three of the most populated campaigns in our \ktwo sample, with both RGB and RC stars. C4 is located close to the Galactic plane and has a median extinction of $A_V \sim 0.8$ mag, while C3 and C5 are in the halo and show much milder extinction ($A_V \sim 0.15$ mag). It appears that, at magnitudes brighter than $G=11.5$ mag, the parallax difference of C4 stars decreases towards more negative offsets with a steeper slope than the other two campaigns. This colour-dependent effect could be an illustration of a reddening systematic (given that C4 contains more red stars with respect to C3 and C5). We also note that the brightest magnitude bin is missing in C4: a lack of bright red stars potentially points towards dust attenuation. The brightest RC stars in the different \ktwo campaigns are similarly distant due to similar spatial distribution. Hence, a RC star in the plane at the same distance as one in the halo would be fainter.

\subsubsection{Is there a need to rescale the extinction?}
\label{sec:syst_rescale}

An error in the extinction correction would more readily affect stars at bright magnitudes than at faint magnitudes. Since Fig. \ref{fig:k2fields} indicates such a trend, we consider whether a small adjustment to the extinction estimate in \texttt{PARAM} can lead to better agreement. We test the impact that a different $R_V$ value would have on the following relationship $A_V = R_V \, E(B-V)$. This implies modifying the asteroseismic parallax as follows:
\begin{align}
	\varpi_{\rm PARAM, new} = \varpi_{\rm PARAM} \cdot 10^{0.2 \, A_{\rm V, PARAM} \, (f - 1 )} \, ,
\end{align}
where $f$ is the correction factor applied to the total extinction in the $V$-band. A factor $f > 1$ is equivalent to $R_V > 3.1$, and $f < 1$ corresponds to $R_V < 3.1$. We also initially considered a correction in the form of an additive term, i.e. $A_V + \delta A_V$, but two significant issues would arise in this case: the corrections derived would often lead to negative extinctions (which are non-physical), and there is no way to tell whether an additive term corrects the extinction, magnitude, or any other quantity involved in the computation of the parallax. Hence, we did not explore this avenue further. Our rescaling test was done using RC stars only.

We first defined a reference low-extinction sample, containing the 30\% stars with the lowest extinctions. Then, for each $G$ magnitude bin (ranging [9,13], [9,15], and [9,11] with a bin size varying between 0.2 and 0.5 mag for \kepler, \ktwo, and \tess, respectively), we derive mean estimates for the asteroseismic and \gaia parallaxes and compute a parallax offset --- this will be our expected value. Then, we keep the 70\% remaining stars as individual. We estimate $\chi^2$ for each $f$ value in the range [0.8, 1.2] as follows:
\begin{align}
	\chi^2 = \left(\frac{\delta}{\sigma}\right)^{2} = \frac{(\Delta \varpi_{\rm ind} - \Delta \varpi_{\rm exp})^2}{\sigma_{\rm \Delta \varpi_{\rm ind}}^2 + \sigma_{\rm \Delta \varpi_{\rm exp}}^2} \, \rm ,
\end{align}
where $\Delta \varpi_{\rm ind}$ and $\Delta \varpi_{\rm exp}$ are the individual and expected parallax offsets after applying the extinction rescaling factor $f$, while $\sigma_{\rm \Delta \varpi_{\rm ind}}$ and $\sigma_{\rm \Delta \varpi_{\rm exp}}$ are their respective uncertainties. All individual $\chi^2$ values are added together so that we have a single $\chi^2$ estimate for a given $f$ and magnitude bin. The best-fitting correction factor $f$ is the one at which the $\chi^2$ distribution finds its minimum, and the corresponding uncertainty is obtained by looking at how much $f$ varies if we were to consider $\min(\chi^2) + 1$.

\begin{figure}
	\centering
	\includegraphics[width=\hsize]{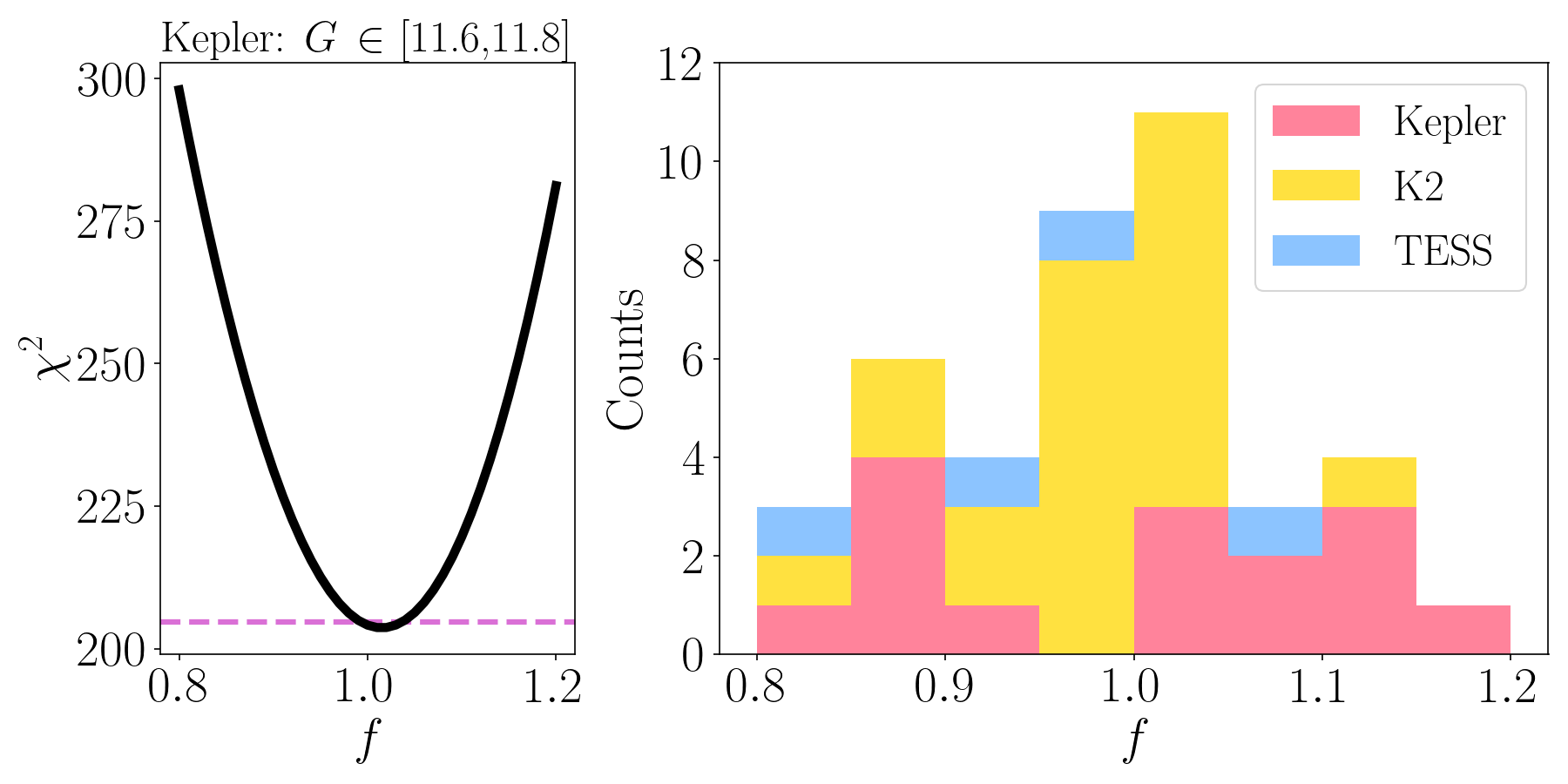}
	\caption{\textit{Left panel:} Example of a $\chi^2$ distribution for a given magnitude bin in the \kepler field. The purple dashed line corresponds to $\min(\chi^2) + 1$. \textit{Right panel:} Stacked distribution of $f$ values in all magnitude bins for \kepler, \ktwo, and \tess.}
	\label{fig:fav}
\end{figure}

Whether it be for \kepler, \ktwo, or \tess, the $\chi^2$ distributions suggest that $f$ is close to unity ($f_{\rm Kepler} \sim 1.01 \pm 0.11$, $f_{\rm \ktwo} \sim 0.99 \pm 0.06$, and $f_{\rm \tess} \sim 0.96 \pm 0.08$ where the two values quoted are the mean and standard deviation) and we do not find any significant correction that should be applied to extinction values from \texttt{PARAM} (Fig. \ref{fig:fav}). Despite this, we know that low-extinction stars are nevertheless more reliable because any uncertainties of the extinction correction will impact the results less, hence we will apply the following criterion for all fields:
\begin{align*}
A_V \leq 0.5 \, \rm mag.
\end{align*}

We also looked at whether we could characterise the sightline extinction, as a function of Galactic latitude. In order to do this, we split our \kepler and \ktwo datasets between the low-extinction reference and other stars, as done above. We look at \ktwo campaigns individually, and do not bin in magnitude as the statistics would be too low. We compute a rescaling factor $f$ for \kepler and each \ktwo campaign with enough low-extinction stars (fields with $|b| < 30^\circ$ are thus excluded), and study the relation with Galactic latitude. We do not observe any clear trend. Considering \kepler and \ktwo campaigns together, we measure a mean value of $f \sim 0.9$, with a standard deviation of about 0.15. We estimate the impact that $f = 0.9$ would have on \kepler and \ktwo parallax differences, restricting the sample to $A_V < 0.5$ mag. We find that the systematic bias due to extinction can be approximated by an exponential with base 10 function:
\begin{align*}
	\sigma_{\rm extinction} \sim 8 \cdot 10^{0.2 \, (11-G)} \; \rm \muas. 
\end{align*}
This relation is valid for $G \in$ [9, 15] mag.

\subsection{Spectroscopic surveys}
\label{sec:syst_spectro}

\texttt{PARAM} uses \teff, \feh, and \afe (when available) from spectroscopy to derive distances, and hence parallaxes. Parameters from different spectroscopic surveys may differ depending on the wavelength in which observations were taken, on the region of the sky they focused on, and also on the spectral resolution. APOGEE DR17 is a near-infrared (NIR) all-sky spectroscopic survey \citep{Abdurrouf2022} with a resolution $R \sim 22 500$, while GALAH DR3 focused on the southern hemisphere in the optical/NIR and $R \sim 28 000$ \citep{Buder2021}. We estimated the systematic uncertainty of the parallaxes by comparing the difference among datasets for the same stars. If we consider \ktwo stars with observations from both catalogues, we are left with nearly 2800 RGB and RC stars. As done in Sect. \ref{sec:syst_astero}, the bias due to the choice of spectroscopic constraints is computed as the absolute mean difference between the offsets measured either using APOGEE or GALAH data. The bias shows small oscillations as a function of magnitude, but overall we find that the spectroscopic systematic uncertainty is about:
\begin{align*}
	\sigma_{\rm spectro} \sim 3 \, \rm \muas .
\end{align*}
We adopted APOGEE DR17 as the largest sample measured homogeneously, across all \kepler, \ktwo, and \tess fields.

\subsection{Photometric data used by \texttt{PARAM}}
\label{sec:syst_phot}

\begin{table*}
	\small
	\caption{Photometric bands used for the different fields.}
	\label{table:bands}
	\centering
	\begin{tabular}{c | c c c c c c c c c c c c c c c c}
		\hline\hline
		Band & $B$ & $g$ & $B_P$ & $V$ & $r$ & $G$ & $i$ & $R_P$ & $z$ & $J$ & $H$ & $K$ & $W1$ & $W2$ & $W3$ & $W4$ \\
		\hline
		$\lambda_{\rm c}$ [$\rm \mu m$] & 0.442 & 0.477 & 0.532 & 0.540 & 0.623 & 0.673 & 0.762 & 0.797 & 0.913 & 1.2 & 1.6 & 2.1 & 3.4 & 4.6 & 12 & 22 \\
		\hline
		\kepler & & X & & & X & & X & & X & X & X & X & X & X & X & X \\
		\hline
		\ktwo & X & X & X & X & X & X & X & X & & X & X & X & & & & \\
		\hline
		\tess-SCVZ & X & & X & X & & X & & X & & X & X & X & X & X & & \\
		\hline
	\end{tabular}
\end{table*}

\begin{figure*}
	\centering
	\includegraphics[width=0.33\hsize]{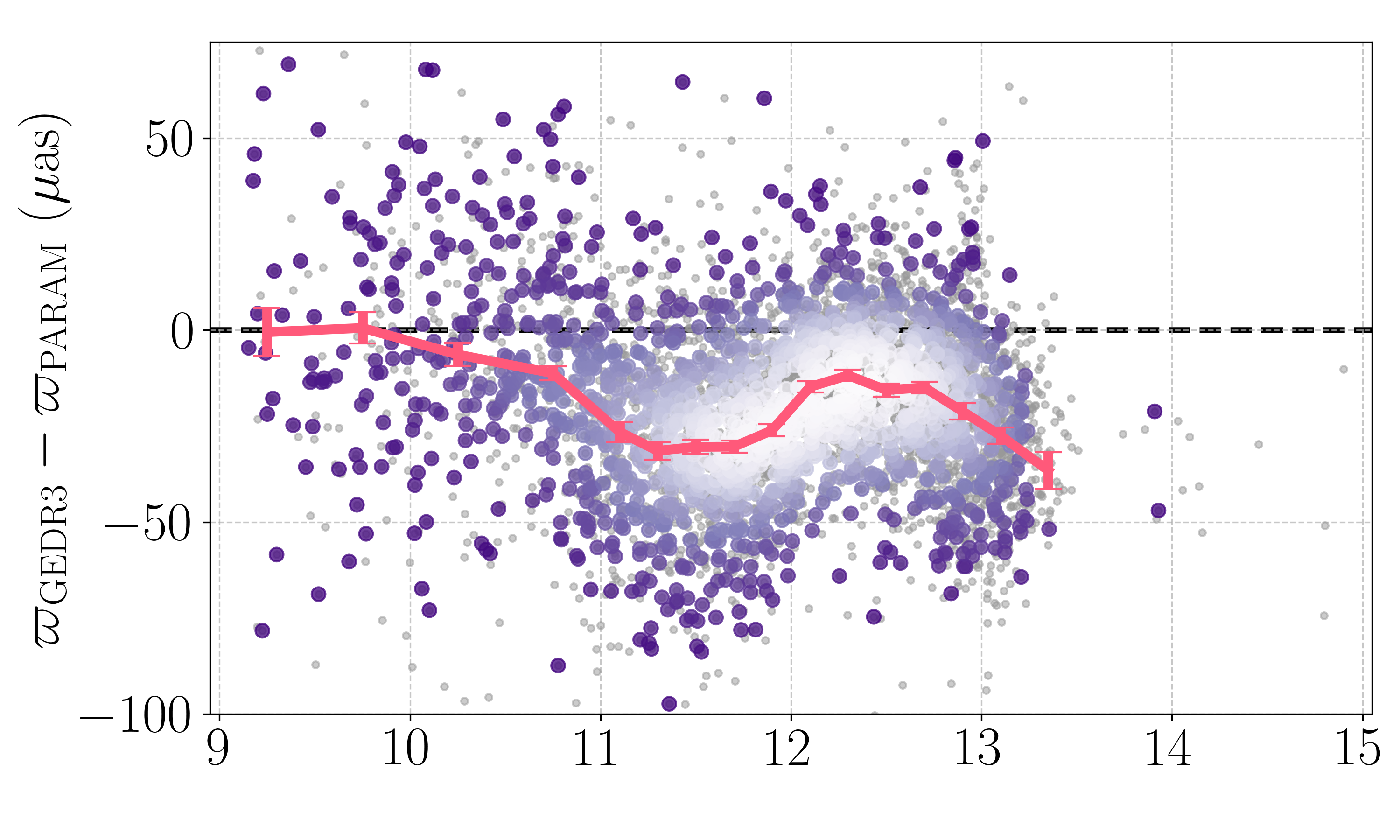}
	\includegraphics[width=0.33\hsize]{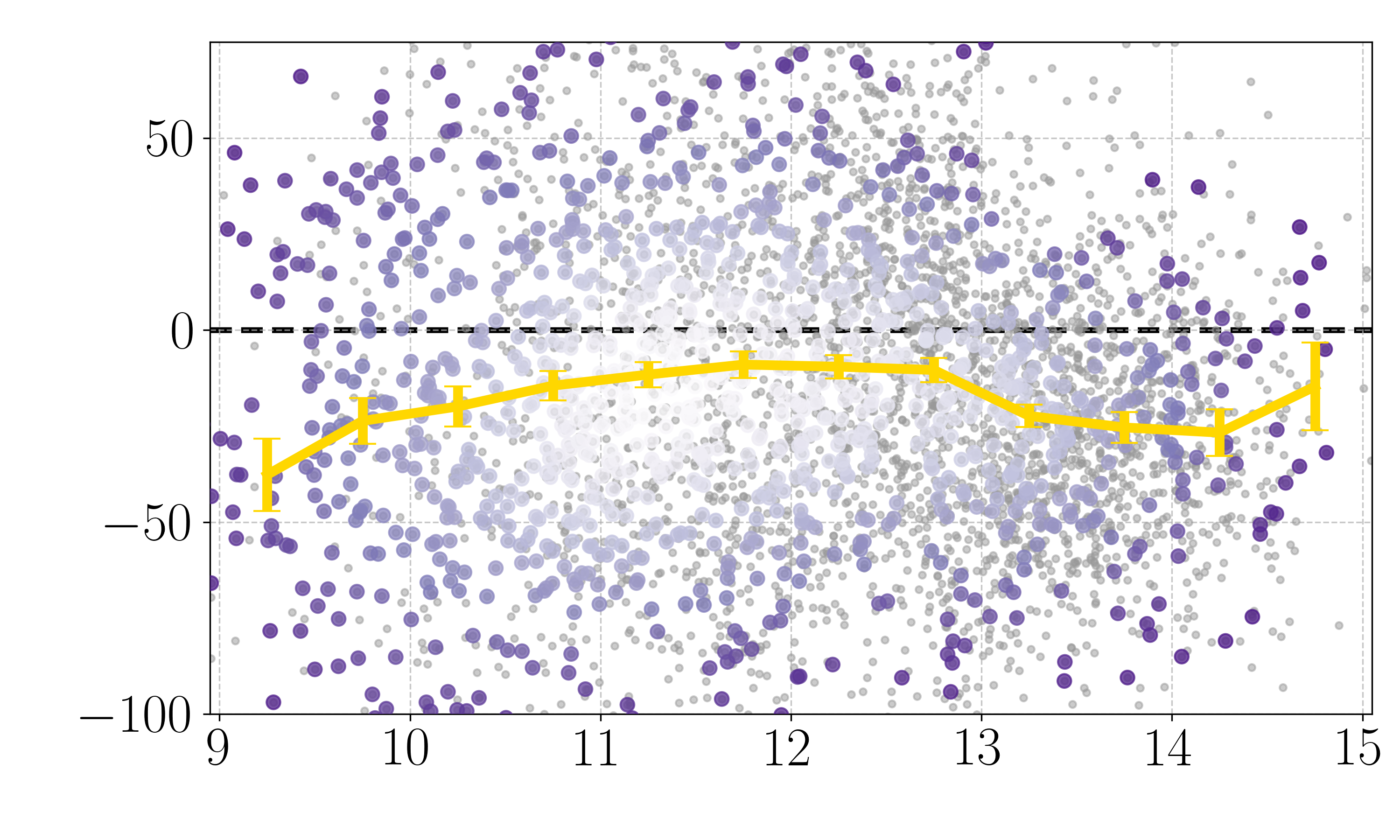}
	\includegraphics[width=0.33\hsize]{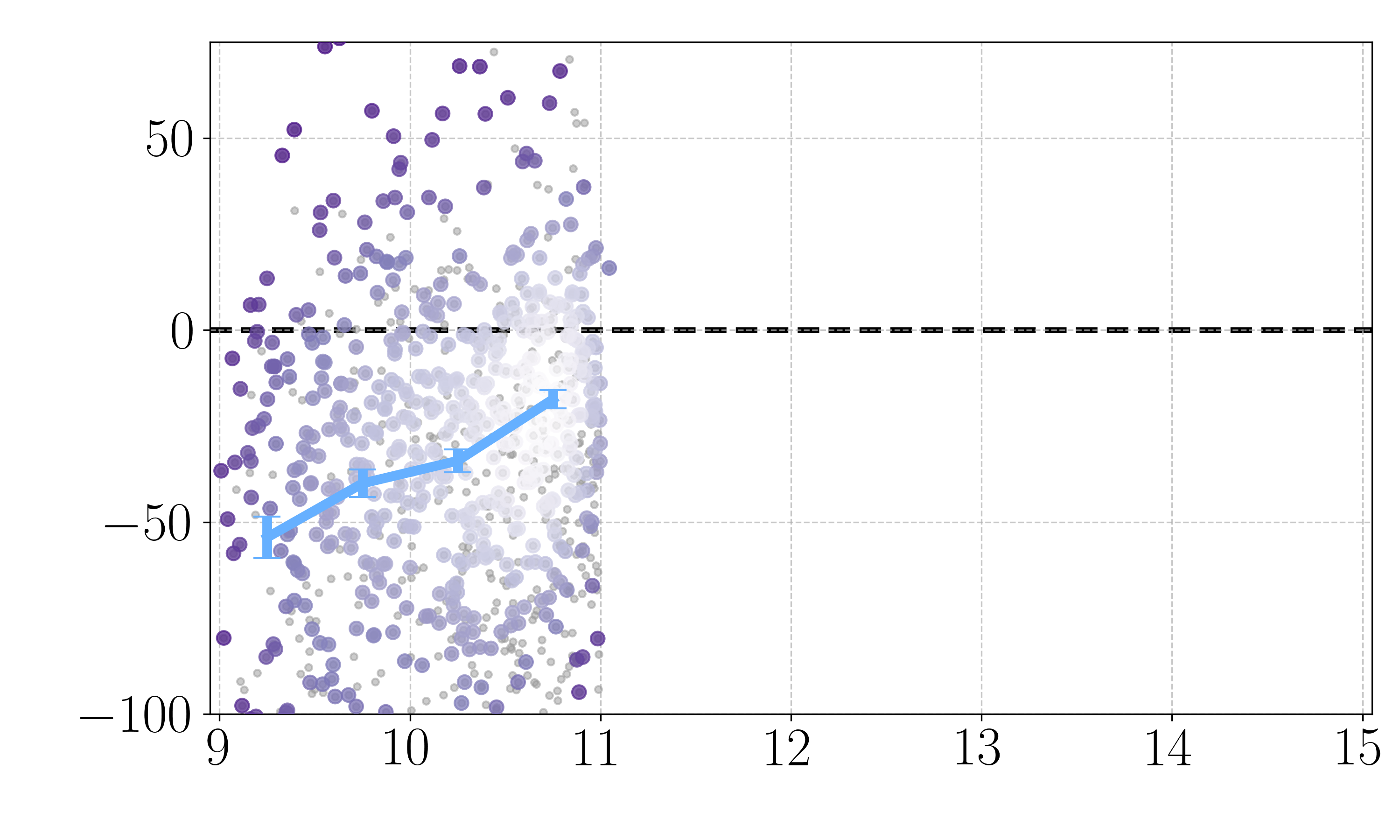}
	\includegraphics[width=0.33\hsize]{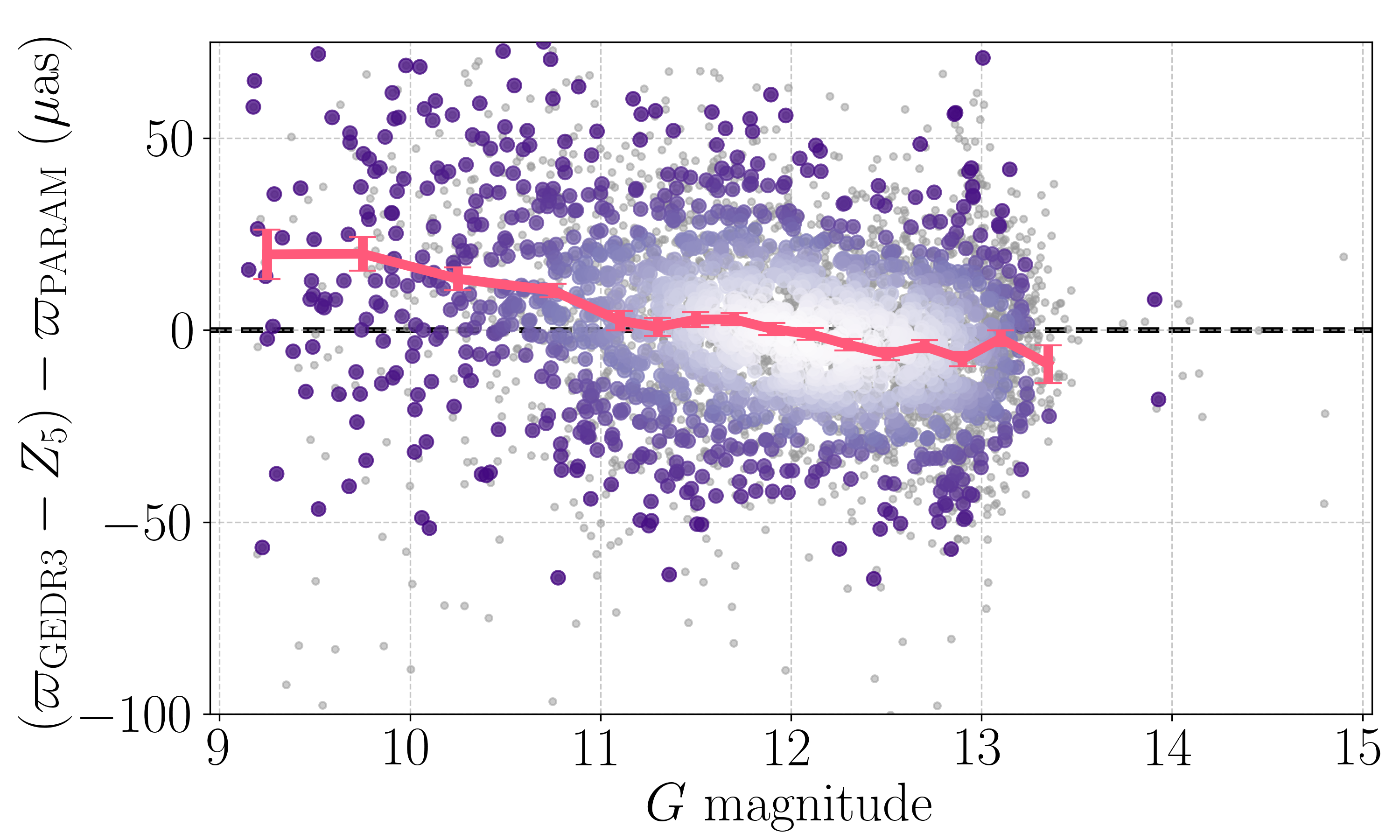}
	\includegraphics[width=0.33\hsize]{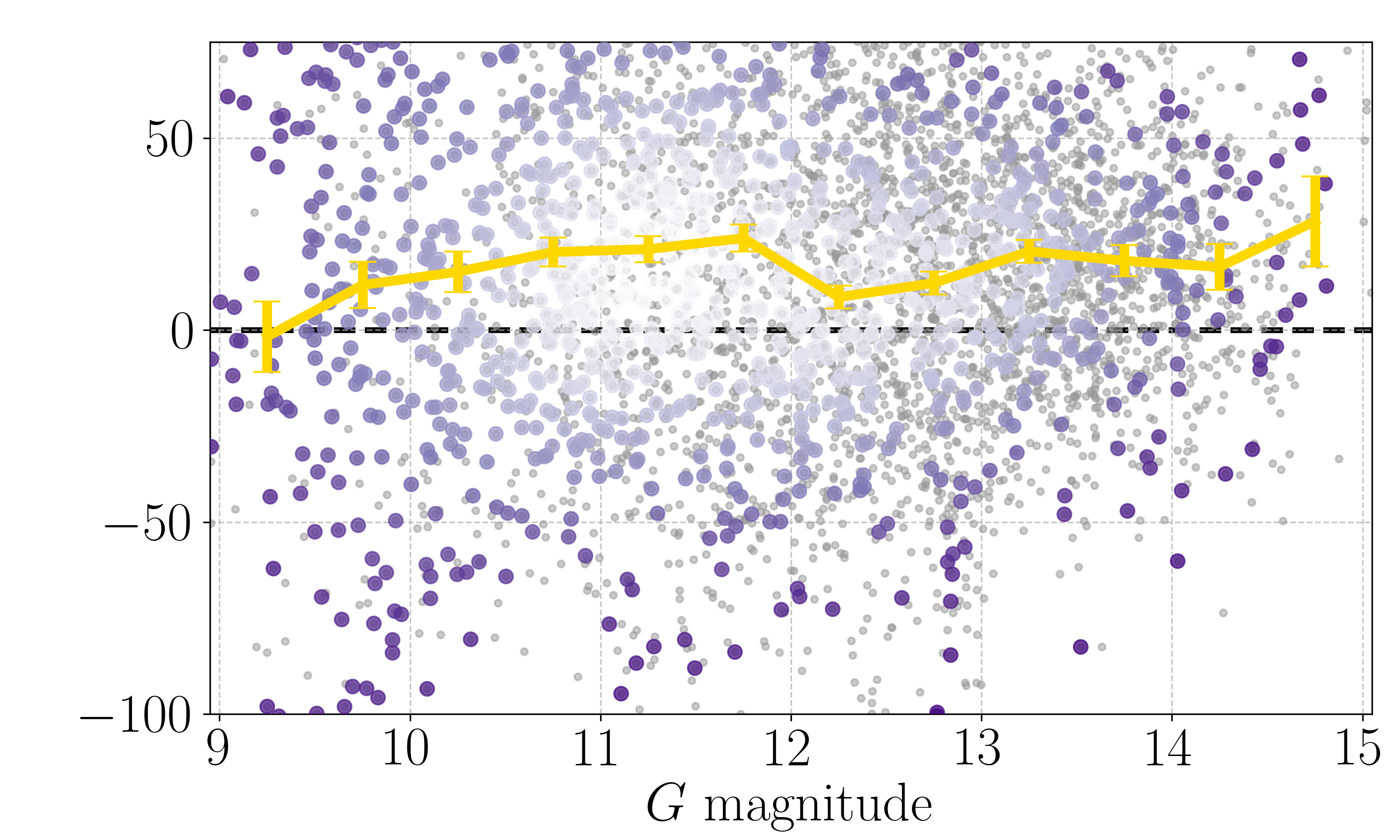}
	\includegraphics[width=0.33\hsize]{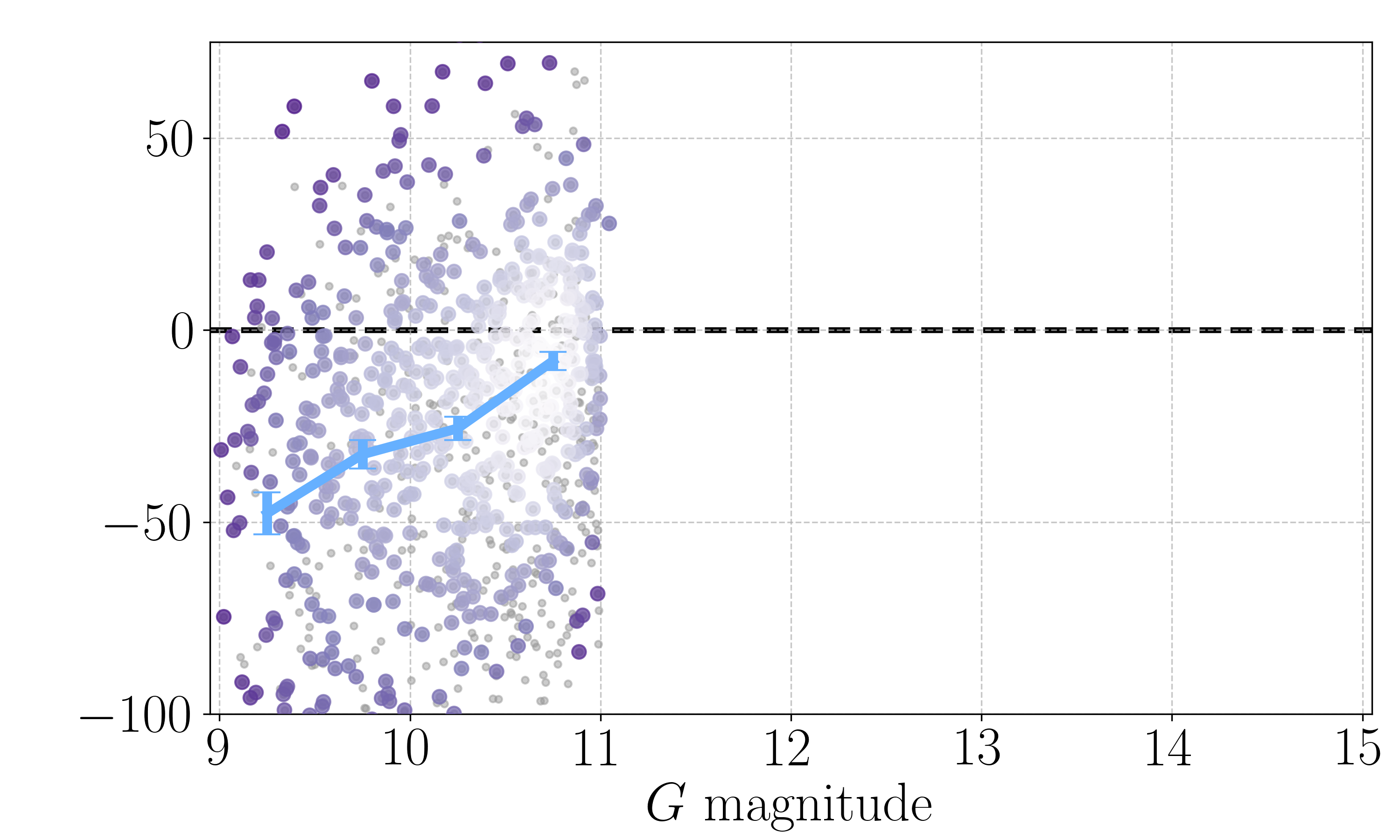}
	\caption{Parallax difference before (top) and after correction (bottom) as a function of $G$ magnitude for \kepler (left), \ktwo (middle), and \tess (right). The grey scatter corresponds to the entire sample of RGB and RC stars. Our best sample with low-extinction RC stars is shown with purple points, with the colour scale indicating density (increasing from dark purple to white). The red, yellow, and blue lines show the running mean and its uncertainty for each field.}
	\label{fig:results}
\end{figure*}

The distances in Paper I used data from different photometric surveys as available. This resulted in different sets of multi-band photometry employed for different fields (\kepler, individual \ktwo campaigns, \tess-SCVZ). Those are explicited in Table \ref{table:bands}, and could lead to slight differences in the homogeneity of the outputs generated by the Bayesian code. In particular, the photometry used for \kepler and \tess goes deeper in the infrared, with respect to \ktwo, notably with the addition of WISE bands.

We also consider the fact that the angular resolution achieved by some of these surveys is not at the level of \gaia's. For instance, it is of the order of $2.5 \rm "$ for 2MASS and $\sim 6 \rm "$ for W1-3. This means that the magnitudes measured in these bands could potentially be affected by blending issues, in the case there would be nearby contaminants not being resolved due to the limited angular resolution.

To test this, we performed a multi-cone search within the \gaia catalogue to list all the sources within a 6" search radius of our targets in \kepler, \ktwo, and \tess. We find that 574, 108, and 106 targets have potential stellar contaminants in each field, respectively. We measured the magnitude contrast between those contaminants and the main source of interest. We consider that blending effects can be considered negligible with a magnitude contrast above 5 mag, which would correspond to a 1/100 flux ratio. We removed stars with at least one nearby source that could contribute significantly to blending, and found them to be 97, 24, and 7 in \kepler, \ktwo, and \tess respectively. If we wanted to decrease the magnitude contrast threshold to 2 mag, it would only remove 6, 0, and 1 star. In both cases, when we repeated our analysis with the reduced samples, the differences were barely noticeable. Hence, potential blending issues are not affecting our results, and we do not have to consider them further in our investigation. \\

Table \ref{table:systematics} provides average estimates for the asteroseismic, extinction, spectroscopic, and total systematic uncertainty in \muas for different $G$-magnitude ranges. We also provide the corresponding systematic error that would apply to the distance modulus, in mag.

\begin{table}
	\caption{Average systematic uncertainty for the asteroseismic, extinction, and spectroscopic components, and the total systematic uncertainty.}
	\label{table:systematics}
	\centering
	\begin{tabular}{c | c c c c}
		\hline\hline
		$G$ range & $\langle \sigma_{\rm seismo} \rangle$ & $\langle \sigma_{\rm extinction} \rangle$ & $\langle \sigma_{\rm spectro} \rangle$ & $\langle \sigma_{\rm total} \rangle$ \\
		\hline
		[9, 11] & 20 \muas & 13 \muas & 3 \muas & 24 \muas \\
		\hline
		[11, 13] & 8 \muas & 5 \muas & 3 \muas & 10 \muas \\
		\hline
		[13, 15] & - & 2 \muas & 3 \muas & 4 \muas \\
		\hline\hline
		all & 0.017 mag & 0.018 mag & 0.006 mag & 0.025 mag \\
		\hline
	\end{tabular}
	\tablefoot{The first three lines give the systematics on parallax in \muas, for various $G$-magnitude ranges (in mag). The last line provides the systematic uncertainty on the distance modulus, in mag. Note that $\sigma_{\rm seismo}$ is not available for the faintest magnitude range, because it has been estimated using \kepler stars which do not go faint enough. $\sigma_{\rm spectro}$ has been found not to vary with $G$, so its value remains constant. The total systematic uncertainty was computed by adding the individual uncertainties in quadrature.}
\end{table}

\section{Analysis of the \gaia EDR3 parallax zero-point}
\label{sec:gaiazp}

\begin{figure}
	\centering
	\includegraphics[width=\hsize]{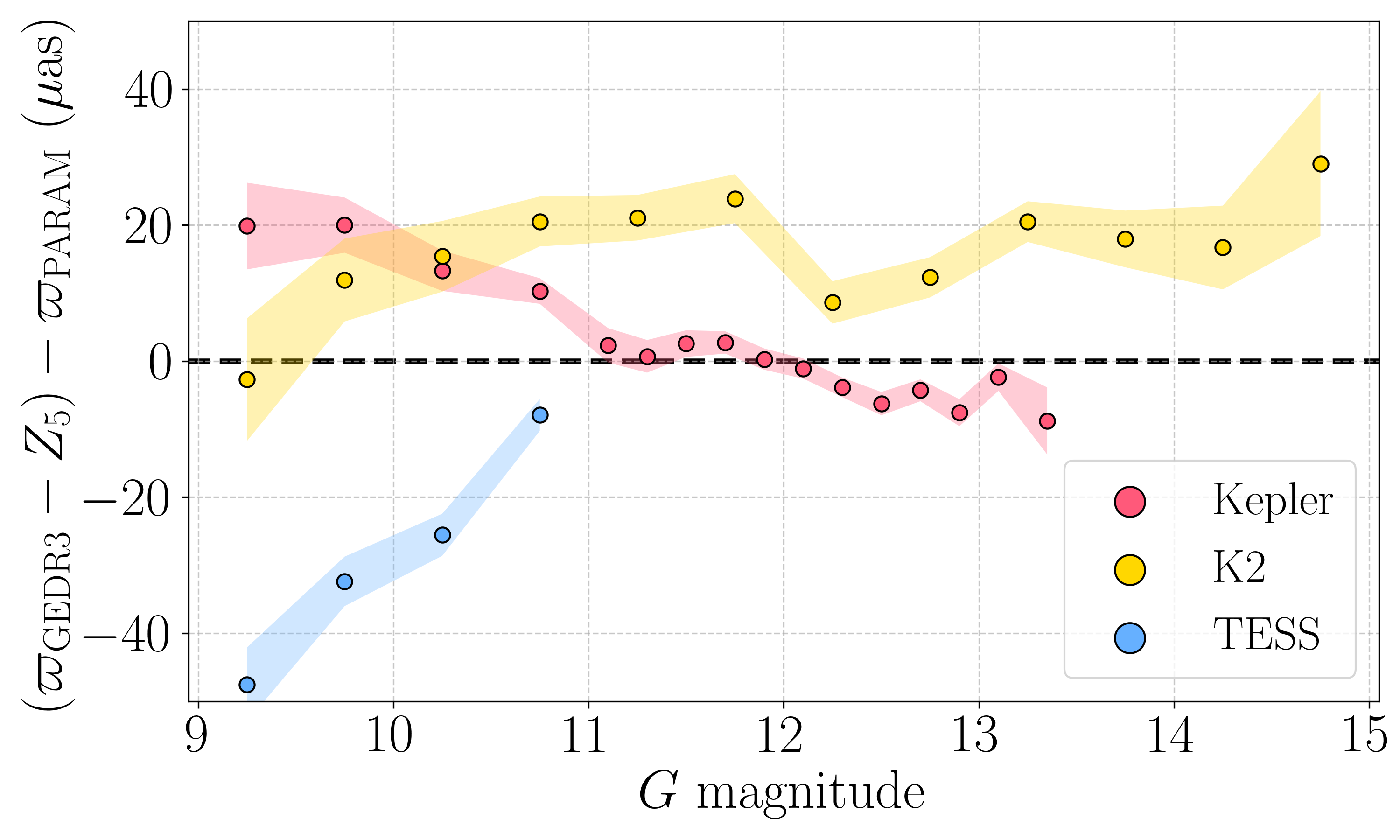}
	\caption{Parallax offset residual (after correcting using \citetalias{Lindegren2021}) as a function of magnitude for \kepler (red), \ktwo (yellow), and \tess (blue). The points show the mean values, while the areas indicate the corresponding uncertainty (given as the error on the mean per bin).}
	\label{fig:residual}
\end{figure}

\begin{figure*}
	\centering
	\includegraphics[width=0.8\hsize]{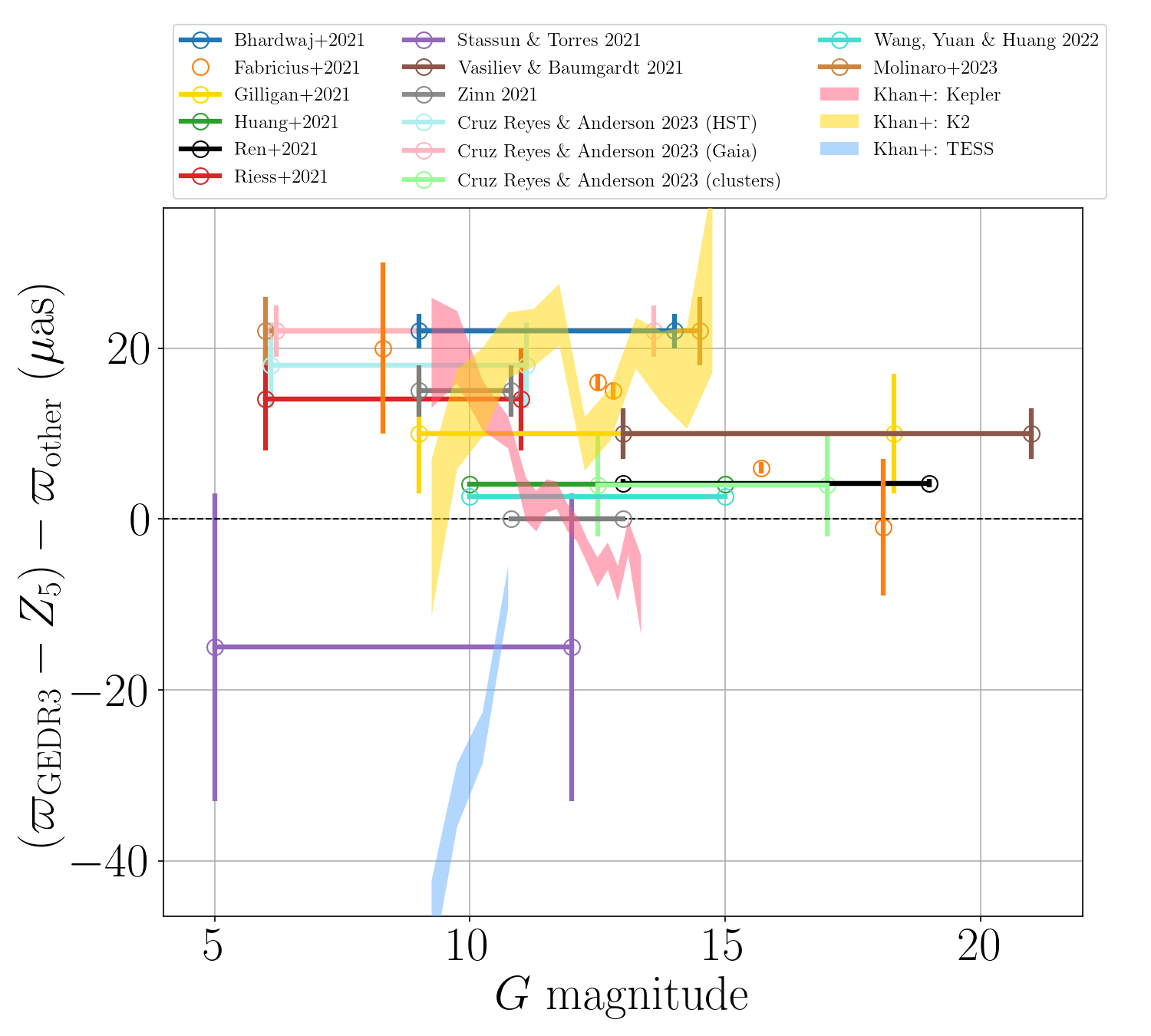}
	\caption{Comparison of our results, $\Delta \varpi_{\rm corr} = (\varpi_{\rm GEDR3}-Z_5) - \varpi_{\rm PARAM}$, to a compilation of \gaia parallax offset residuals from the literature, as a function of the magnitude in the $G$-band. Results displayed as open circles with vertical error bars are from the literature. Our findings for \kepler, \ktwo, and \tess are shown as red, yellow, and blue filled areas, respectively. Different literature sources used astrophysically very different objects and methods. Here we use the same methodology applied to an intrinsically rather homogeneous group of stars. The sign convention used here is such that positive values correspond to an overcorrection of parallaxes, and negative values to an underestimation of parallaxes by \citetalias{Lindegren2021}.}
	\label{fig:compilation}
\end{figure*}

As described above, we selected the following sample to obtain the most detailed view of \gaia's parallax systematics using asteroseismic parallaxes:
\begin{itemize}
	\item red clump stars selected according to temperature, iron abundance, and absolute magnitude (cf. Fig. \ref{fig:hrd});
	\item asteroseismic constraints from \citetalias{Elsworth2020}, as the pipeline gives more consistent results between RGB and RC stars and we do not expect the parallax offset to depend on the evolutionary stage (see Fig. \ref{fig:astero});
	\item spectroscopic properties from APOGEE DR17;
	\item low-extinction stars ($A_V \leq 0.5$ mag);
	\item and \gaia EDR3 sources with good astrometric quality (RUWE $< 1.4$).
\end{itemize}
With these choices, we are left with 1560, 1227, and 635 RC stars in \kepler, \ktwo, and \tess, respectively. \\

The top panels of Fig. \ref{fig:results} show $\Delta \varpi = \varpi_{\rm EDR3} - \varpi_{\rm PARAM}$ as a function of magnitude for \kepler, \ktwo, and \tess RC targets. Background stars in grey show the entire sample with both RGB and RC stars, while the purple ones correspond to RC stars only. The running mean is calculated based on RC stars and shows more evidently the trend with magnitude. The non-linear trend with $G$ is very much apparent in \kepler, as already discussed in Sect. 4.1 of Paper I. We do not see it in \ktwo possibly because of the larger uncertainties or the fact that \ktwo fields are scattered along the ecliptic which might average out some systematics, while \tess targets are too bright. Overall, $\Delta \varpi$ values are mostly negative, but they do approach zero, e.g. for $G \leq 10$ mag in \kepler, and at $G \sim 12.5$ mag for \ktwo.

The bottom panels of Fig. \ref{fig:results} show instead $\Delta \varpi_{\rm corr} = (\varpi_{\rm EDR3}-Z_5) - \varpi_{\rm PARAM}$ as a function of magnitude. For \kepler stars, the magnitude-dependent non-linear systematics are very effectively removed by the \citetalias{Lindegren2021} corrections. The parallax offset residuals become very close to zero for $G \geq 11$ mag. \ktwo just seems to be shifted globally towards slightly positive residuals, i.e. \citetalias{Lindegren2021} corrections are a bit too large. And \tess is shifted by a very small amount, which is far from being enough to reach zero, i.e. in that case \citetalias{Lindegren2021} offsets are underestimated.

Figure \ref{fig:residual} gathers the residuals obtained for \kepler, \ktwo, and \tess together, as a function of magnitude. At bright magnitudes ($G \leq 11$ mag), the picture is not that clear. For \kepler, \citetalias{Lindegren2021} overcorrects the \gaia parallaxes. It is also the case for \ktwo fields, despite a decrease towards zero at the brightest end. This goes in the same direction as the literature results in this magnitude regime \citep[see, e.g.,][]{Li2022,Molinaro2023}. However, \tess exhibits the opposite trend, i.e. \citetalias{Lindegren2021} undercorrects the parallaxes. $Z_5$ is of the order of $\sim 10 \, \rm \muas$ for \tess, and $\sim 25 \, \rm \muas$ for \kepler. Given the similarity of the stars and the homogeneous methodology, we consider a dependence on sky location in Sect. \ref{sec:k2camp}. But whether it be for \kepler or \tess, it either corrects too much or not enough, so this potentially points out a drawback in the \citetalias{Lindegren2021} zero-point correction model. There also does not seem to be a link between the trend suggested by Fig. \ref{fig:residual}, and the fact that we have varying observation lengths in \kepler (4 years), \tess-SCVZ (1 year), and finally \ktwo (80 days), in descending order (see Sect. \ref{sec:kepvsk2} for a discussion about how observing conditions may affect the target selection and asteroseismic properties). All of this means that one cannot simply apply the same offset at bright magnitudes.

Figure \ref{fig:compilation} shows our results in the context of other recent findings related to \gaia parallax systematics. These literature offsets were determined in different regions of the sky using astrophysically different objects and different methods. There is no clear trend with magnitude, although the majority of residuals are positive. A summary of the median parallax offsets and residuals obtained for each field in the bright and faint magnitude ranges is given in Table \ref{table:results}. \\

We also considered the possibility to apply additional \gaia photometric quality flags, in order to remove stars likely to be blended by nearby sources \citep[see Sect. 9.3 in][]{Riello2021}. These are:
\begin{itemize}
	\item $\beta = (\texttt{phot\_bp\_n\_blended\_transits}+\texttt{phot\_rp\_n\_blended\_transits})*1.0/(\texttt{phot\_bp\_n\_obs}+\texttt{phot\_rp\_n\_obs}) < 0.1$ (blending fraction);
	\item \texttt{ipd\_frac\_multi\_peak} $< 7$ (\gaia EDR3 percent of successful-IPD windows with more than one peak);
	\item \texttt{ipd\_frac\_odd\_win} $< 7$ (\gaia EDR3 percent of transits with truncated windows or multiple gates);
	\item and $C^{*} < 1 \sigma$ (\gaia EDR3 corrected BP/RP excess factor).
\end{itemize}
The numbers of stars in each field then become 1314, 1051, and 474. This affects the median values computed at the level of $\pm 2 \, \muas$ at most, so it is not significant.

\begin{table}
	\caption{Median parallax offsets before and after applying \citetalias{Lindegren2021}, in the faint ($11 \leq G \leq 13$ mag) and bright magnitude ranges ($G \leq 10$ mag).}
	\label{table:results}
	\centering
	\begin{tabular}{c | c c c}
		\hline\hline
		Fields & $G$ range & $\langle \Delta \varpi \rangle$ ($\rm \mu as$) & $\langle \Delta \varpi_{\rm corr} \rangle$ ($\rm \mu as$) \\
		\hline
		\kepler & [11, 13] & $-21.5$ & $-1.6$ \\
		& $\leq 10$ & $+0.2$ & $+19.7$ \\
		\hline
		\ktwo & [11, 13] & $-10.1$ & $+16.5$ \\
		& $\leq 10$ & $-27.5$ & $+7.4$ \\
		\hline
		\tess-SCVZ & $\leq 10$ & $-44.9$  & $-37.6$ \\
		\hline
	\end{tabular}
\end{table}

\section{Discussion}
\label{sec:discussion}

\subsection{How to interpret differences between \kepler and \ktwo}
\label{sec:kepvsk2}

At faint magnitudes ($G > 11$ mag), \kepler and \ktwo lead to different conclusions regarding the suitability of the \citetalias{Lindegren2021} zero-point model. The latter works well for \kepler targets, where the residual offset is almost zero. However, for \ktwo fields, $Z_5$ overcorrects \gaia parallaxes.

Even if the asteroseismic pipeline leading to the measurement of \numax and \deltanu is the same for both \kepler and \ktwo, the two surveys still have significant differences observation-wise. \kepler is a high ecliptic latitude field with observations taken continuously for 4 years. \ktwo campaigns cover 17 different locations (in our study), and their observations span a much shorter duration of 80 days. So their comparison should account for the larger uncertainties in \ktwo, and we cannot interpret our results in the sense that \ktwo contradicts \kepler.

The difference in the observation duration leads to differences in the seismic performance index \citep{Mosser2019}. This index depends on \numax, the height-to-background ratio, and the observation duration. It is naturally lower for shorter observations, as in \ktwo. Hence, measurements are likely to be biased towards red-giant stars with a high enough signal/amplitude. For a given magnitude, the signal ratio between observations spanning either several years or a couple of months can reach 10 or more \citep[see, e.g., Fig. 3 in][]{Mosser2019}. Population effects, associated with different magnitude distributions for instance, can also affect the seismic signal. Fainter targets, which \ktwo contains in larger fraction, are increasingly harder to detect asteroseismically. We note that our selection of RC stars in \ktwo (cf. bottom middle panel of Fig. \ref{fig:hrd}) effectively removes metal-poor RHB stars that could be biased in their \deltanu, hence radius and mass, determination \citep{Tailo2022,Matteuzzi2023}. As a sanity check, we also exclude $\alpha$-rich stars ([$\alpha$/Fe] > 0.15 dex) from our \ktwo RC sample and this brings the residual offset to $\sim +14$ \muas (for $G \in$ [11, 13] mag), hence nothing significantly different.

All these aspects go in the same direction, such that \kepler's observing conditions are more favourable to the detection of oscillations and to a greater quality of asteroseismic measurements. However, \ktwo also provides invaluable information regarding spatial variations of the parallax zero-point, that could go beyond the only latitudinal dependence considered in \citetalias{Lindegren2021}, as we will discuss in the next section. The shorter baseline of \ktwo compared to \kepler results in a much reduced density (number/sky area) of red-giant stars with determined asteroseismic parameters. Despite the fact that \ktwo goes fainter and has $\sim 15$ times the total area of \kepler thanks to the many different campaigns, the number of stars in the \kepler and \ktwo samples are similar. This means that the K2 stars could be a biased selection of the population that make up the Kepler sample. Hence, there is a small chance that some of the difference between \kepler and \ktwo comes from population differences that cannot be conclusively assessed here. Yet, a comparison between the different \ktwo campaigns is valid because the conditions are the same for each campaign (baseline, area, magnitude range).

\subsection{Exploring variations among \ktwo campaigns}
\label{sec:k2camp}

\begin{figure}
	\centering
	\includegraphics[width=\hsize]{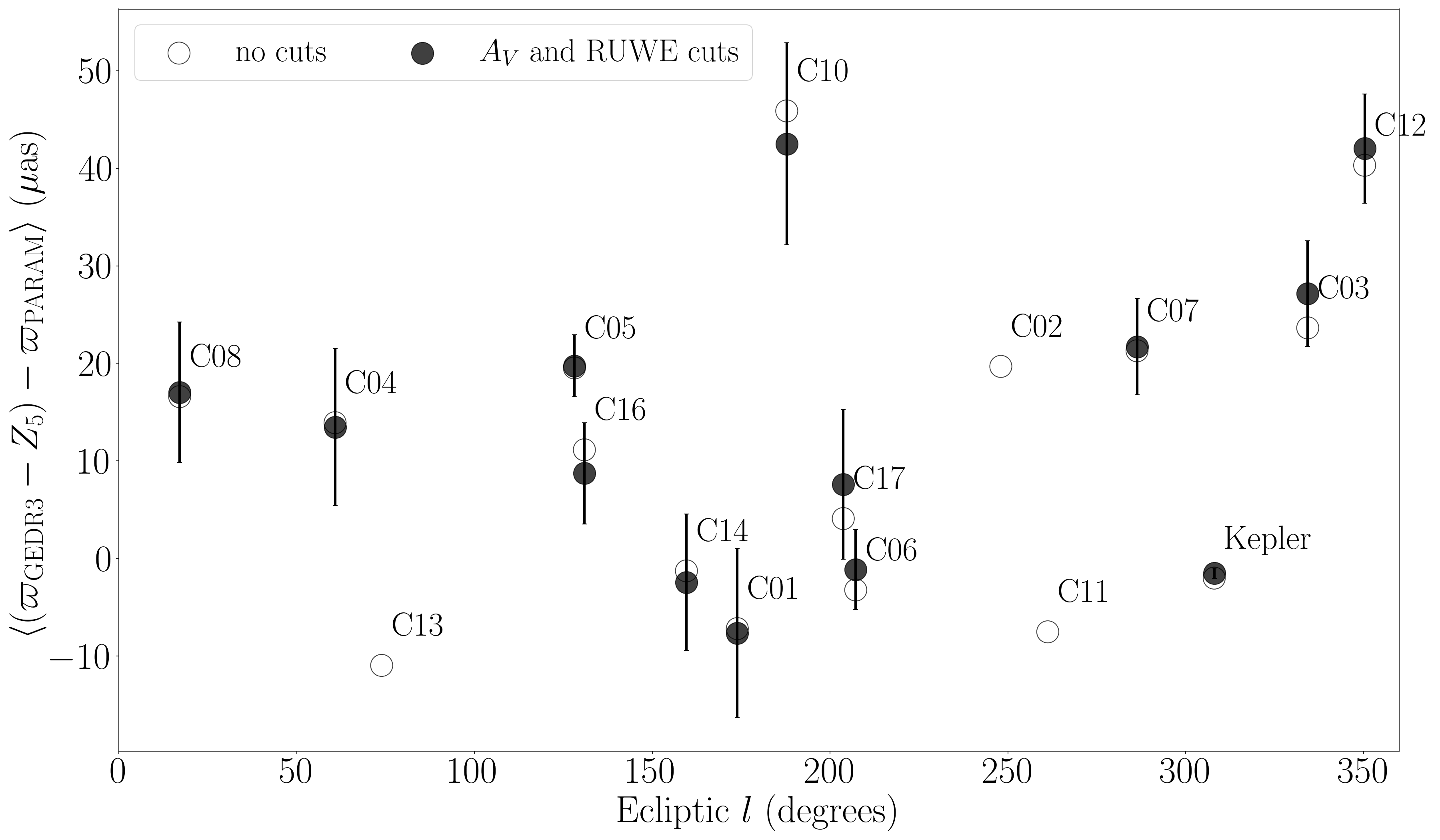}
	\caption{Median residual offset as a function of ecliptic longitude for the individual \ktwo campaigns and \kepler. Open and filled circles show the results before and after applying the extinction and RUWE cuts. For clarity, error bars are only shown for the offsets with the cuts applied.}
	\label{fig:k2_ecl_lon}
\end{figure}

\ktwo's survey design probes different ecliptic longitudes at zero ecliptic latitude. This allows to investigate a dependence of the residuals on ecliptic longitude as shown in Fig. \ref{fig:k2_ecl_lon}. To this end, we evaluated the median parallax offset for each \ktwo campaign in the magnitude range $G \in [11, 13]$ mag. We also added the \kepler field for comparison. Campaigns with similar longitude taken years apart yield similar results, which excludes significant temporal evolution of the instrument as a source of the variations. There are some intriguing features that we highlight: 1) most campaigns are near $+10$ or $+20$ \muas, 2) adjacent campaigns (in $l_{\rm ecl}$) can differ greatly (cf. near $180^{\circ}$), 3) maximum variations are seen near $180^{\circ}$ and $360^{\circ}$, 4) there are campaigns at near-identical $l_{\rm ecl}$ (C6 and C17, C5 and C16). Since the analysis is done exactly in the same way using astrophysically very similar objects, we consider the variations real. We see that there are two clear outliers with a residual offset $\gtrsim +40 \rm \; \muas$: C10 at $l_{\rm ecl} \sim 180^{\circ}$ and C12 at $l_{\rm ecl} \sim 360^{\circ}$. All the other campaigns lie below $\sim +30 \rm \; \muas$. Including the extinction and RUWE flags slightly decreases the overall scatter among the different \ktwo campaigns, as one would expect since that restricts the sample to stars with better photometry and astrometry. 

Hence, \ktwo campaigns provide a spatial dependence hint, here along the ecliptic longitude --- which is currently not considered in the \citetalias{Lindegren2021} correction model. When combining all \ktwo fields together, this longitudinal dependence is averaged over, and data from very different regions of the sky are combined, which is not the case for the original \kepler field and \tess-SCVZ. This stresses the point that analyses must still solve the residual parallax offset and cannot simply adopt offsets from the literature if these were not determined on similar samples with respect to sky location, magnitude, and colour. So it is likely that the $\sim +16 \, \rm \muas$ ($+13 \, \rm \muas$, when the two greatest outliers C10 and C12 are removed) residual offset observed in the $G \in [11,13]$ mag range could be caused by systematic uncertainties related both to the quality of asteroseismic observations (Sect. \ref{sec:kepvsk2}) and the need to resolve positional variations within \ktwo (which requires higher statistics).

\subsection{Coverage of bright physical pairs in \citetalias{Lindegren2021} model}
\label{sec:l21coverage}

\begin{figure*}
	\centering
	\includegraphics[width=\hsize]{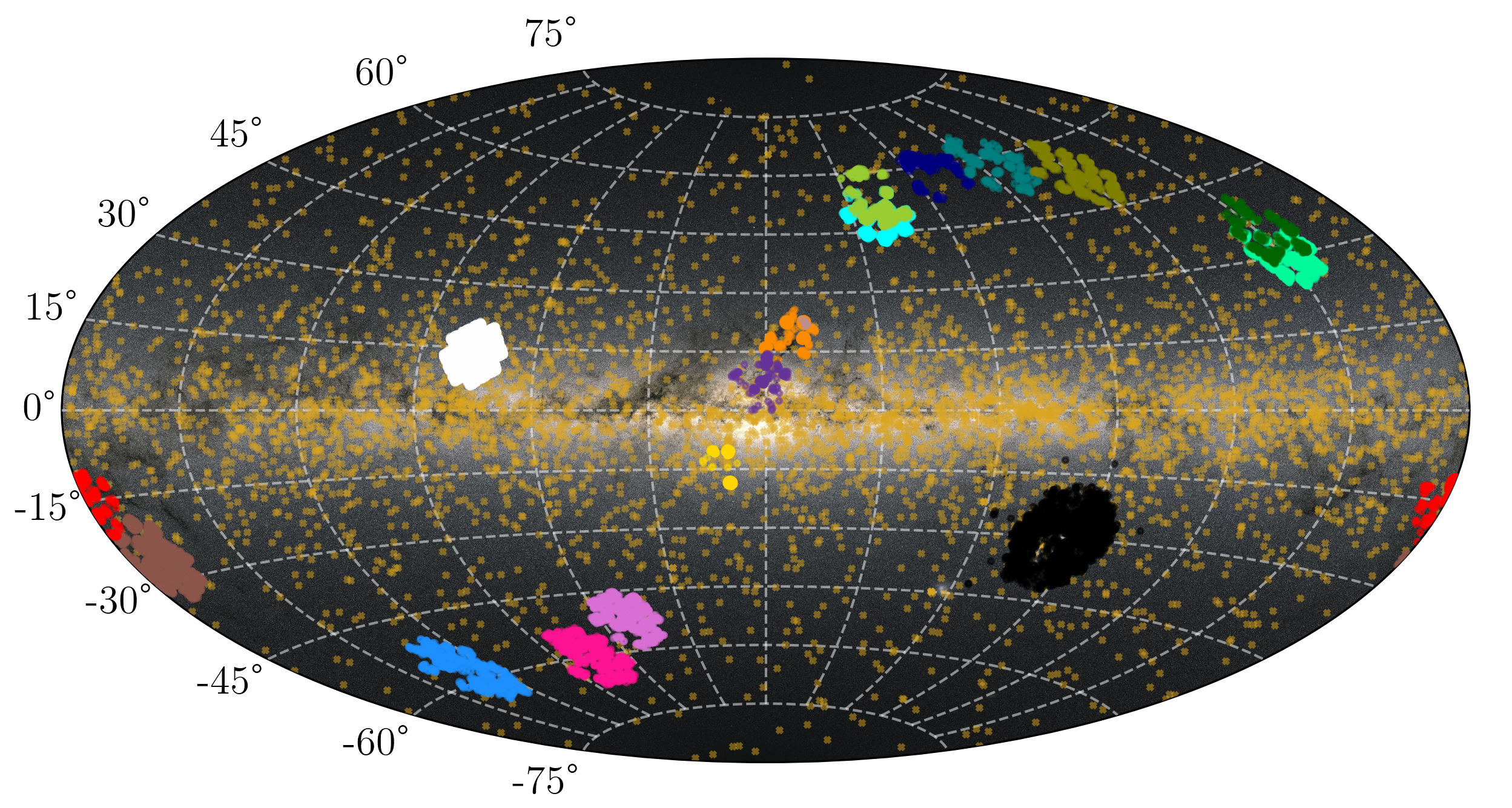}
	\includegraphics[width=0.25\hsize]{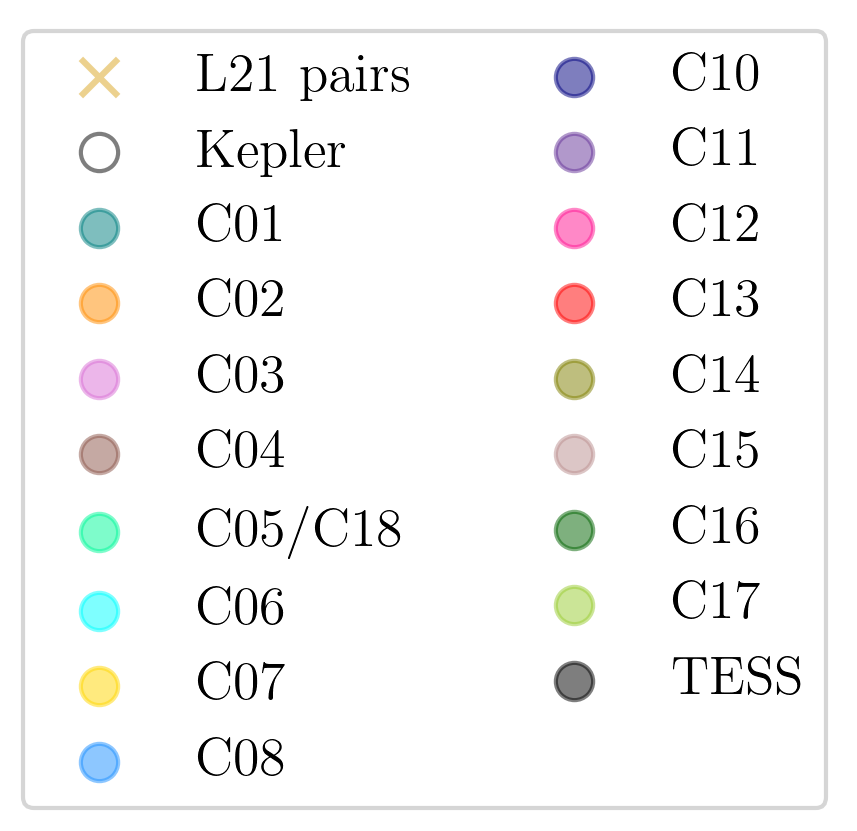}
	\caption{Skymap in Galactic coordinates, showing the location and coverage resulting from the crossmatch between the various asteroseismic fields considered in this study and APOGEE DR17. Yellow crosses correspond to the location of the bright physical pairs ($G < 11$ mag) that have been used in the process of deriving the zero-point correction model from \citetalias{Lindegren2021}. This figure has been generated using the \texttt{python} package \texttt{mw-plot} (\url{milkyway-plot.readthedocs.io}). The background image comes from ESA/Gaia/DPAC.}
	\label{fig:skymap}
\end{figure*}

The \citetalias{Lindegren2021} zero-point correction is based on quasars, physical pairs, and stars in the LMC. What is missing is thus a good determination of this bias for brighter stars: quasars only cover the faint end of magnitudes ($G > 14$ mag), very few LMC sources have $G < 13$ mag (1457), and physical pairs constitute the major part of calibrators in the bright regime ($6 < G < 14$ mag) with nearly 70,000 sources but only $\sim 4000$ with $G < 11$ mag. Hence, the bright component of the correction relies quite heavily on the sky coverage by physical pairs. We reproduced Fig. 1 from \citet{Khan2023}, overlaying on top the location of the physical pairs with $G < 11$ mag that have been used within the \citetalias{Lindegren2021} correction (list of sources shared through priv. comm. by Lindegren; see Fig. \ref{fig:skymap}). Among the initial list of nearly 121,000 physical pairs, only 4374 are brighter than $G=11$ mag. The coverage provided by these bright physical pairs is rather sparse over the entire sky: most of them are located on the Galactic disc, whilst the asteroseismic fields probe many off-disc locations. Given the limited overlap and low statistics at the bright end, we consider our reported differences to indicate true shortcomings of the \citetalias{Lindegren2021} model that originate from the limited spatial coverage by physical pairs. This is also demonstrating how asteroseismology can effectively complement \gaia in certain regions of the sky.

\subsection{Estimation of the red clump magnitude}

\begin{figure}
	\centering
	\includegraphics[width=\hsize]{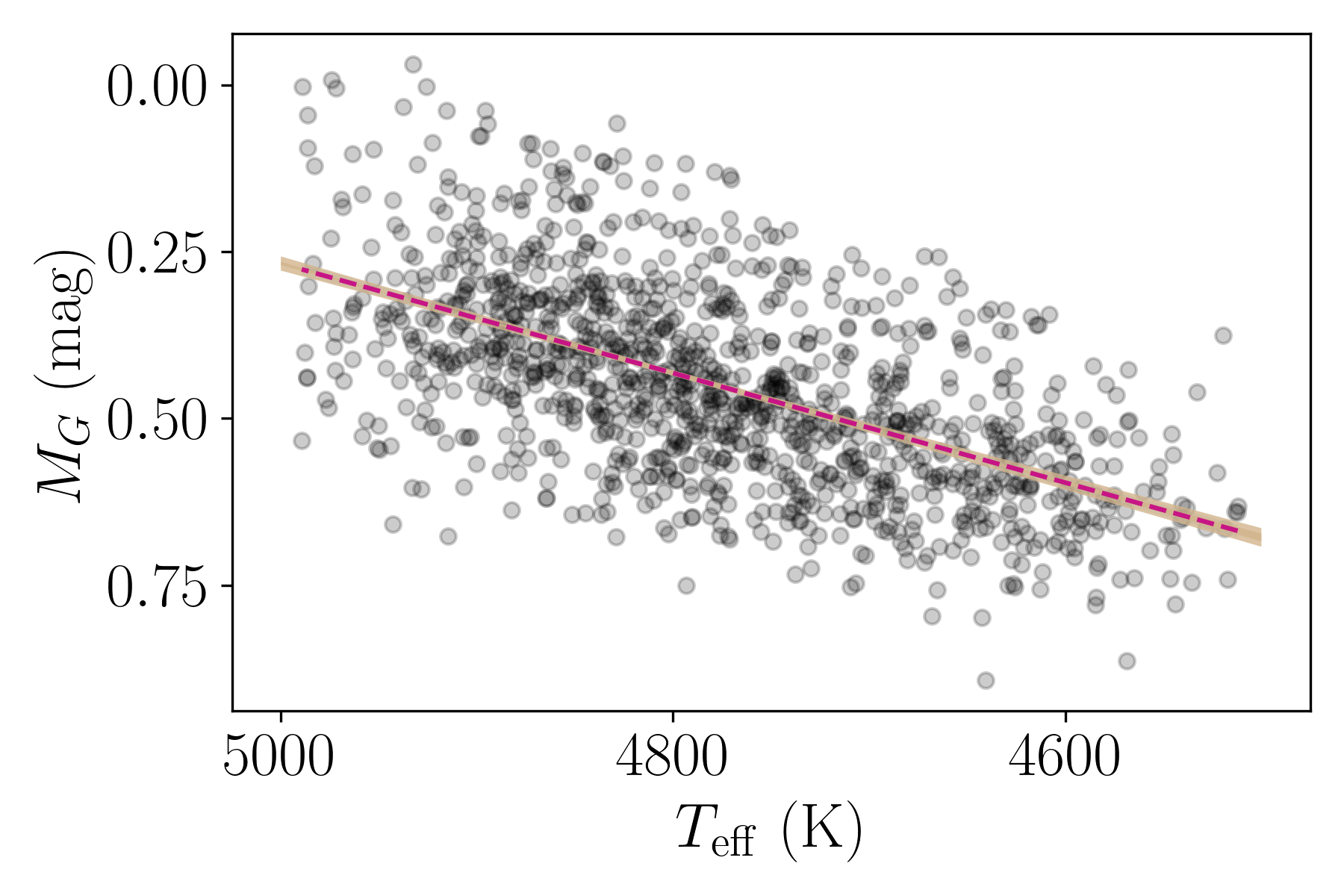}
	\caption{Hertzsprung-Russell Diagram for our final set of \kepler RC stars where the $G$-band absolute magnitude is computed using \citetalias{Lindegren2021}-corrected \gaia parallaxes (black points). The sample is restricted to $G \in [11, 13]$ mag, where we found the residual parallax offset to be $\sim$ zero. The pink dashed line shows the fit given by Eq. \ref{eq:MG}, and the brown shaded region illustrates how the fit could vary given the uncertainties on the slope and the intercept.}
	\label{fig:kepler_rcmag}
\end{figure}

In Sect. \ref{sec:gaiazp}, we find a robust result for the \kepler field, namely that the residual parallax offset (after applying \citetalias{Lindegren2021} corrections) is nearly null for $11 < G < 13$ mag. We use this opportunity to compute the absolute magnitude of the red clump in this magnitude range, in the $K_s$ and $G$ bands. Absolute magnitudes and uncertainties are computed for individual RC stars as follows:
\begin{align}
	M_{\lambda} = m_{\lambda} + 5 \log_{10} (\varpi_{\rm EDR3} - Z_5) + 5 - A_{\lambda} \, ,\\
	\sigma_{M_{\lambda}} = \sqrt{\sigma_{m_{\lambda}}^2 + ((5/\ln10) \times (\sigma_{\rm \varpi_{\rm EDR3}}/\varpi_{\rm EDR3}))^2 + \sigma_{\rm A_{\lambda}}^2} \, ,
\end{align}
where $\varpi_{\rm EDR3}$ and $Z_5$ are in arcseconds, and $A_{\lambda}$ is calculated from the $A_V$ estimate computed by \texttt{PARAM}. Taking the median value, we find the absolute magnitude of the clump to be 
\begin{align}
	M_{K_s}^{\rm RC} & = -1.650 \pm 0.025 \, \rm mag
\end{align}
in the $K_s$ band. The uncertainties are computed in a conservative way, such that we consider the total systematic uncertainty from Table \ref{table:systematics} (0.025 mag) and the formal error on the median (0.002-0.003 mag). The latter being fairly negligible, the systematic error completely dominates in this case. Unlike $M_{K_s}^{\rm RC}$, which is relatively independent of \teff, $M_G^{\rm RC}$ can vary quite significantly over the \teff range considered here. Hence, we instead state the absolute magnitude in the $G$-band as a function of \teff:
\begin{align}
	\begin{split}
	M_{G}^{\rm RC} & = (0.432 \pm 0.004) \\
	               & - (0.821 \pm 0.033) \cdot (\teff [\rm{K}] - 4800\rm{K})/1000\rm{K} \, \rm [mag], 
	\end{split}
	\label{eq:MG}
\end{align}
where the average dispersion around the fit is of $\sim 0.12$ mag. Figure \ref{fig:kepler_rcmag} shows the HRD for the final sample of \kepler stars used to estimate $M_G^{\rm RC}$, and the fit derived as a function of \teff (Eq. \ref{eq:MG}). As discussed in Sect. 5.1 of Paper I, the RC magnitude is also subject to population effects related to the age and metallicity spread so the conservative estimate of the uncertainty reflects that intrinsic scatter as well. Literature values for the $K_s$ band had been found to range between $-1.53$ and $-1.62$ mag before the \gaia era \citep[see Table 1 in][]{Girardi2016}. With \gaia DR1, \citet{Hawkins2017} derived $M_{G}^{\rm RC} = 0.44 \pm 0.01$ mag and $M_{K_s}^{\rm RC} = -1.61 \pm 0.01$ mag. Using \gaia DR2 parallaxes, \citet{Hall2019} measured $M_{G}^{\rm RC} = 0.546 \pm 0.016$ mag and $M_{K_s}^{\rm RC} = -1.634 \pm 0.018$ mag, while \citet{Chan2020} found $M_{G}^{\rm RC} = 0.435 \pm 0.004$ mag and $M_{K_s}^{\rm RC} = -1.622 \pm 0.004$ mag.

Our RC absolute magnitude estimates are in reasonably good agreement with the literature, within the uncertainties. For the $G$-band RC magnitude, our estimate agrees with \citet{Hawkins2017} and \citet{Chan2020} for $\teff \sim 4800$ K, while a cooler \teff (4650 K) is required to obtain a good agreement with results from \citet{Hall2019}. 
We acknowledge that the RC absolute magnitude determination presented here is slightly circular because the same RC stars that determine the absolute magnitude were used to demonstrate the absence of a residual zero-point in the magnitude range used to determine $M_{\lambda}^{\rm RC}$. Additionally, the extinction estimates are obtained by \texttt{PARAM} and rely on known SEDs and bolometric corrections. However, because the parallax offset residual is zero and extinction corrections are low by selection, it is still valid to present this despite the circularity.

\section{Conclusions}
\label{sec:conclusions}

This study provides the most in-depth asteroseismic view of \gaia parallax systematics based on the most informative subset of stars in our sample, namely RC stars observed by \kepler, \ktwo, and \tess-SCVZ, together with a detailed analysis of uncertainties affecting the asteroseismic parallaxes. It builds upon the asteroseismic datasets and methods presented in Paper I \citep{Khan2023}. 

We first discuss the important role of red clump stars as standard candles, and the benefits we gain by focusing on them, instead of using mixed datasets containing both RGB and RC stars. These benefits are the following: RC stars have a $\sim$ fixed luminosity rendering any trends with apparent magnitude easier to interpret, they are less affected by asteroseismic radius/distance biases, they are a large and more homogeneous stellar population alone.

We then discuss and quantify uncertainties, notably of systematics, related to red clump stars' asteroseismic parallaxes. These include differences among asteroseismic methods, extinction, spectroscopy, and photometry. The contribution of each of these is provided in Table \ref{table:systematics}, either as parallax systematics in \muas in given $G$ magnitude ranges, or as distance modulus systematics in mag. These tests allow us to define the dataset considered `best' for analysing parallax systematics. This is constituted of red clump stars, with asteroseismology from \citetalias{Elsworth2020}, spectroscopy from APOGEE DR17, low extinctions ($A_V \leq 0.5$ mag), and RUWE $< 1.4$.

We present detailed results for parallax offsets as a function of magnitude in \kepler, \ktwo, and \tess-SCVZ fields, both before and after applying \citetalias{Lindegren2021} corrections. Each field leads to different conclusions regarding the parallax systematics residuals. For magnitudes fainter than $G=11$ mag, the residuals are very close to zero in \kepler. However, in the bright regime ($G \leq 11$ mag), the \citetalias{Lindegren2021} correction model overcorrects parallaxes for \kepler targets by $\sim +15 \, \rm \muas$, and undercorrects those of \tess sources by $\sim -25 \, \rm \muas$ on average. For \ktwo fields, \citetalias{Lindegren2021} offsets are slightly overestimated independently of magnitude (by $\sim +12 \, \rm \muas$). Previous literature compilations have led to the general finding that parallax residuals follow a trend with $G$ magnitude, meaning that the \citetalias{Lindegren2021} model significantly overcorrects parallaxes for $G \leq 13$ mag and that this effect would decrease on the fainter side \citep{Li2022,Molinaro2023}. Our findings show that the final picture is actually not as straightforward. In the same bright magnitude regime ($G \leq 11$ mag), \kepler, \ktwo, and \tess either demonstrate that the \citetalias{Lindegren2021} model overestimates or underestimates the parallax systematics. This conclusion is found using a dataset as homogeneous as possible, with the same asteroseismic observables and spectroscopic constraints (\citetalias{Elsworth2020} and APOGEE DR17). All three fields share similar magnitude and colour ranges. We also made sure to impose the same \teff and \feh range, when selecting RC stars. The main difference lies in their location on the sky. Hence, it could be that the \citetalias{Lindegren2021} offsets do not fully account for the positional dependence of the systematics. Moreover, studies should not apply magnitude-dependent residual parallax offset corrections, nor rely on residual parallax offsets derived from objects covering different parameter spaces (e.g. sky location, magnitude). This is particularly true for objects brighter than $G = 11$ mag. For the time being, studies seeking parallax accuracy better than 15 \muas should solve for the residual parallax offset as it applies to the sample under study.

Lastly, we discuss observational differences that exist between \kepler and \ktwo, and that one needs to consider when interpreting our results. Population differences between the samples could play a role and should be further considered. However, comparisons among \ktwo campaigns are valid because they are derived from very similar data. We emphasise that our findings for \kepler are the most reliable ones. We also explore \ktwo campaigns in more detail, by checking individual parallax offsets against ecliptic longitude. We potentially find interesting variations, where two campaigns stand out with significantly more positive offsets at $l_{\rm ecl} \sim 180^{\circ}$ and $360^{\circ}$. The $\sim +16 \, \muas$ residual offset in the faint magnitude regime ($G \in [11,13]$ mag), where \kepler agrees very well with \citetalias{Lindegren2021}, could also in part be related to this longitudinal dependence not being resolved when considering the \ktwo dataset as a whole. There is certainly room for improvement as \tess observations will potentially pursue for another decade, the ESA PLATO mission will observe thousands of solar-like oscillators in a field of view $\sim$ 20 times larger than \kepler and is expected to launch by the end of 2026 \citep{Rauer2014}, and HAYDN (one of five potential candidates for the next ESA medium-class mission) would offer the possibility to probe much denser fields, such as globular clusters, with a potential launch in the late 2030s \citep{Miglio2021a}. We discuss the suitability of the \citetalias{Lindegren2021} model for bright targets. We consider the differences at bright magnitudes to indicate real shortcomings of the \citetalias{Lindegren2021} model, which is based on rather sparse physical pairs at $G < 11$ mag. And finally, we compute the absolute magnitude of the red clump in the $K_S$ and $G$ bands, using \kepler RC stars with $11 < G < 13$ mag for which the residual offset is basically zero. These are the first measurements using \gaia EDR3 parallaxes, and they are in close agreement with previous results from the literature within the uncertainties.

\nocite{Bhardwaj2021}
\nocite{Fabricius2021}
\nocite{Gilligan2021}
\nocite{Huang2021}
\nocite{Ren2021}
\nocite{Riess2021}
\nocite{Stassun2021}
\nocite{Vasiliev2021}
\nocite{Zinn2021}
\nocite{CruzReyes2023}
\nocite{Wang2022}

\begin{acknowledgements}
We warmly thank Lennart Lindegren for providing information regarding the \gaia parallax zero-point correction model. We also wish to thank the referee whose comments helped clarify and improve the paper. This work has made use of data from the European Space Agency (ESA) mission
{\it Gaia} (\url{https://www.cosmos.esa.int/gaia}), processed by the {\it Gaia} Data Processing and Analysis Consortium (DPAC, \url{https://www.cosmos.esa.int/web/gaia/dpac/consortium}). Funding for the DPAC has been provided by national institutions, in particular the institutions participating in the {\it Gaia} Multilateral Agreement. RIA and SK are funded by the Swiss National Science Foundation (SNSF) through an Eccellenza Professorial Fellowship
(award PCEFP2\_194638). AM acknowledges support from the ERC Consolidator Grant funding scheme ({\em project ASTEROCHRONOMETRY}, G.A. n. 772293). This research was supported by the International Space Science Institute (ISSI) in Bern, through ISSI International Team project \#490,  SHoT: The Stellar Path to the Ho Tension in the Gaia, TESS, LSST and JWST Era.
\end{acknowledgements}

\bibliographystyle{aa} % style aa.bst
\bibliography{references} % your references Yourfile.bib

\end{document}